\documentclass[11pt]{elsart}
\usepackage{psfig}

\newcommand{\bd}[1]{ \mbox{\boldmath $#1$}  }
\newcommand{\gtrsim} {  \stackrel{\mbox{\large $_>$}}
                                 {\mbox{\large $_\sim$}}  }
\begin{document}

\setlength{\parindent}{1.5em}
\begin{frontmatter}
\title{Comprehensive calculations of three--body breakup cross sections}
\author{E. Garrido}
\address{Instituto de Estructura de la Materia, CSIC, Serrano 123, E-28006
Madrid, Spain}
\author{D.V.~Fedorov and A.S.~Jensen}
\address{Institute of Physics and Astronomy,
Aarhus University, DK-8000 Aarhus C, Denmark}

\date{\today}
\maketitle

\begin{abstract} 
We present in detail a theoretical model for fragmentation reactions
of three--body halo nuclei. The different reaction mechanisms
corresponding to the different processes are described and discussed.
Coulomb and nuclear interactions are simultaneously included and the
method is therefore applicable for any target, light, intermediate and
heavy.  Absolute values of many differential cross sections are then
available as function of beam energy and target.  We apply the method
to fragmentation of $^6$He and $^{11}$Li on C, Cu and Pb. A large
variety of observables, cross sections and momentum distributions, are
computed. In almost all cases we obtain good agreement with the
available experimental data.
\end{abstract}

\end{frontmatter}

\par\leavevmode\hbox {\it PACS:\ } 25.60.-t, 25.10.+s, 21.45.+v

\section{Introduction}
The peculiar structure of halo nuclei is clearly revealed by
fragmentation reactions \cite{rii94,han95,tan96,jon98}.  Large
interaction cross sections and narrow momentum distributions of the
fragments observed after fragmentation of these nuclei are direct
signs of their large spatial extension.  The main properties of these
nuclei are well described by few--body models \cite{zhu93,nie00},
where the nucleus is viewed as an inert core surrounded by a few
(usually one or two) bound nucleons. Among them, Borromean
two--neutron halo nuclei have attracted a lot of attention, and their
most prominent examples, $^6$He and $^{11}$Li, have been widely
investigated.
 
During the last decade a large amount of experimental data after
fragmentation of $^6$He and $^{11}$Li on different targets have been
provided.  In particular total dissociation cross sections and total
interaction cross sections, as well as momentum distribution of the
fragments after dissociation and after core breakup are available
\cite{tan88,kob89,kob92,bla93,nil95,hum95,orr95,zin97,chu97,ale98,aum99,sim98}.
Different models describing these data have been developed. These
models fall in two independent main groups according to the kind of
reaction for which they were designed. The first one considers nuclear
breakup processes, and therefore they apply to the case of light
targets. Most of them describe neutron dissociation reactions while
core breakup is usually omitted \cite{ber98,gar99}.  Only calculations
performed in the framework of the Glauber theory are also providing
total interaction cross sections \cite{oga92,suz94,alk96}. The second
group focuses on heavy targets where only Coulomb dissociation is
considered \cite{suz90,ber92,bar98,cob98}.  Intermediate targets where
nuclear and Coulomb interaction compete is outside the scope of these
models. At best Coulomb and nuclear breakup are computed in
independent models and subsequently simply added \cite{ber93,bar93}.

The reaction mechanisms responsible for the breakup processes are
still controversial \cite{gar00a,gar00b,ers00}, but absolutely
essential to understand and analyze the data properly
\cite{gar00c}. It is established that few-body aspects of the halos
are necessary.  
In most investigations halo constituents (and targets) are
treated as inert particles where the intrinsic degrees of freedom are
neglected or accounted for by finite range potentials with Pauli
forbidden states. However, the core degrees of freedom are
unavoidable for some observables like interaction cross sections where
core destruction is substantial. Such many-body aspects must be included
at some point while maintaining the necessary halo features. 

More than one reaction mechanism is
necessary in a correct description of halo breakup. The long range
Coulomb and the short range nuclear potentials already lead to
clear-cut differences \cite{gar00a,gar00b}. However the finite
extension of halo particles and targets furthermore produces
inevitable differences in the active reaction mechanisms at small and
large impact parameters. This has strong implications for the
understanding and the analyses.

The starting point for developing our model was the recognition that,
for large beam energies ($\gtrsim 100$ MeV/nucleon),
the time scales allow application of the sudden approximation for
two-neutron \cite{gar96,gar97b} and one-neutron halos
\cite{han96}. Only light targets, relative momentum distributions and
the specifically dominating processes were accessible within the
sudden approximation.  The next step was to use a
participant-spectator separation and employ the phenomenological
optical model to describe the interactions between participants and
target \cite{gar98a}. Again only light targets were accessible but now
absolute cross sections were computed. The model was then extended to
treat the interactions between spectators and target in the black
sphere model \cite{gar99}. Finally the Coulomb and nuclear
interactions were included on the same footing allowing treatment of
light, intermediate and heavy targets and computation of absolute
values of all differential breakup cross sections
\cite{gar00a,gar00b}.

The reaction mechanisms now come out as clearly distinguishable for
light and heavy targets or rather for small and large impact
parameters. The small impact parameters involve violent collisions and
one or more of the halo particles do not arrive in the detectors in
the forward direction. At large impact parameters the momentum
transfer is relatively small and the three halo particles are gently
excited into the continuum state, which subsequently decays and all
halo particles most likely arrive in the forward detectors. These two
extremes are separated roughly at impact parameters slightly larger
than corresponding to grazing collisions.  These different reaction
mechanisms imply that the two-step processes, where one halo particle
is removed by the target while the resonance of the remaining system
is populated followed by decay \cite{zin97,chu97,ale98,aum99,sim98},
at best only is consistent with small momentum transfer \cite{gar00c}.
The large momentum transfer requires a completely different mechanism
of halo particle removal.

In this paper we formulate for the first time in detail the model
accounting for all these effects. So far no other model contains all
the necessary features. The idea is to describe the fragmentation
reaction as the sum of all the possible reactions where one, two
or all three halo constituents interact simultaneously with the
target. The processes leading to different final states clearly add
up incoherently. The cross sections for the remaining
processes are in all the numerical applications in this report also
simply added as appropriate for large momentum transfer from the
target. The model is consistently describing two--neutron halo
fragmentation on any target with simultaneous treatment of nuclear and
Coulomb interactions.  Absolute values are computed for all possible
dissociation cross sections distinguished according to the particles
left in the final state. Two--neutron removal cross sections, core
breakup cross sections, interaction cross sections and momentum
distributions are computed as function of target and beam energy above
50 MeV/nucleon.  We shall present results for a large variety of these
observables arising from fragmentation reactions of $^6$He and
$^{11}$Li on C, Cu, and Pb.  These three targets have been chosen as
typical examples for light, intermediate and heavy targets displaying
different reaction mechanisms due to the interplay of Coulomb and
nuclear interactions.  Details of the model are given in
Sections~\ref{sec2} and~\ref{sec3}.  Results for fragmentation
reactions of $^6$He and $^{11}$Li on C, Cu and Pb are shown in
Section~\ref{sec4}. Finally Section~\ref{sec5} contains summary and
conclusions.

\section{Model and method for inert particles}
\label{sec2}

We consider a spatially extended three--body halo nucleus colliding
with a target at high energy. Let us first assume that the target and
the three halo constituents in the projectile are inert particles.
The differential cross section $d\sigma$ is to a good approximation
given by the sum of three terms $d\sigma^{(i)}$, each of them
describing the independent contribution to the process from the
interaction between the target and the halo particle $i$.  This is the
assumption used in the classical formulation for a weakly bound
projectile \cite{ban67}. We neglect the binding energy of the initial
three-body bound state compared to the high energy of the beam.
 
\begin{figure}
\centerline{\psfig{figure=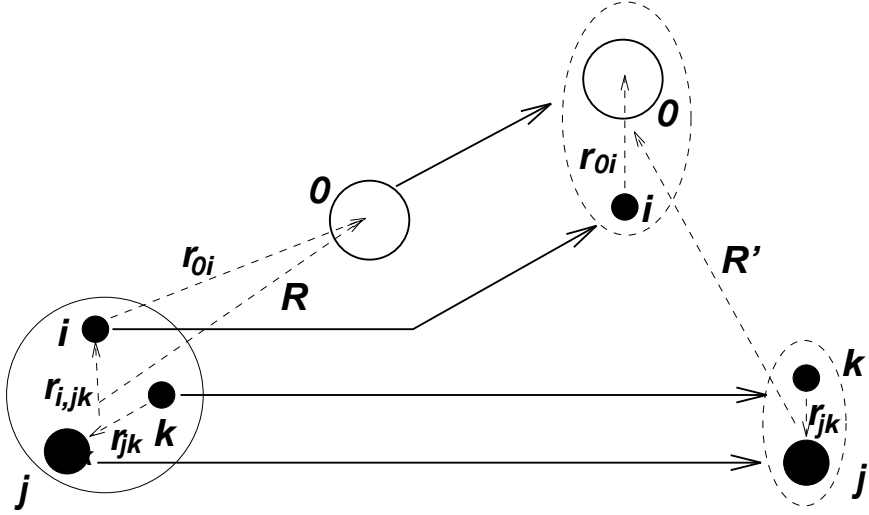,width=19cm,%
bbllx=6cm,bblly=2.cm,bburx=22.4cm,bbury=26cm,angle=270}}
\vspace*{-5cm}
\caption[]{ Sketch of the reaction and coordinates used for the final
state two by two separation. The target is labeled by 0 and
$\{i,j,k\}$ label the particles within the three--body projectile.}
\label{fig1}
\end{figure}

The reaction is then described as three particles independently
interacting with the target as if each particle were free. Each
individual interaction with the target is viewed as removal of one
particle $i$ (participant) while the other two particles $j$ and $k$
(spectators) both survive the reaction undisturbed (see
Fig.~\ref{fig1}).  The participant is either absorbed (in the optical
model sense) or elastically scattered (diffracted) by the target.

The masses, coordinates and conjugate momenta are denoted $m$,
$\bd{r}$ and $\bd{p}$, respectively. The three halo particles and the
target are labeled by $\{i,j,k\}=\{1,2,3\}$ and $0$, respectively.
The relative coordinates $\bd{r}_{jk}$, $\bd{r}_{0i}$,
$\bd{r}_{i,jk}$, $\bd{R}$ and $\bd{R}^{\prime}$ are defined as shown
in Fig.~\ref{fig1}.  The corresponding conjugate momenta are
analogously denoted by $\bd{p}_{jk}$, $\bd{p}_{0i}$, $\bd{p}_{i,jk}$,
$\bd{P}\equiv\bd{p}_{0,ijk}$ and
$\bd{P}^\prime\equiv\bd{p}_{0i,jk}^\prime$. We use primes to denote
the momenta in the final state.  Detailed expressions for the
different relative coordinates and momenta can be found in
\cite{gar99}. The momentum transfer in the reaction for elastic
scattering of the participant is then given by
\begin{equation} \label{mtra}
\bd{q}=\bd{p}_0-\bd{p}^\prime_0=\bd{p}_{0i}-\bd{p}^\prime_{0i} \;\; , \;\;
 q \equiv |\bd{q}| = 2 p_{0i} \sin{\frac{\theta}{2}} \; ,
\end{equation}
where $\theta$ is the angle between $\bd{p}_{0i}$ and $\bd{p}^\prime_{0i}$.

\subsection{Two by two separation in the final state}

For relatively large momentum transfer between participant and target
the final state is appropriately described as two independent
subsystems, i.e. the target plus participant and the two spectators.
This separation is not a matter of convenience but dictated by
physics, especially in connection with the generalization in the next
section to spatially extended particles.  
This is a central assumption reflecting the reaction mechanism. It is
discussed in previous publications and tested by comparison with
experimental results \cite{hum95,zin97,ale98,aum99,gar00c,gar96}.
The transition amplitude of the process in Fig.~\ref{fig1} can then 
within the mixed plane/distorted wave Born approximation be written as
\cite{gar99,sat83}
\begin{equation}
T^{(i)}= \langle \phi_{ \bd{p}_{0i}^\prime  \Sigma^\prime_i}^{(0i-)} 
\phi_{ \bd{p}_{jk}^\prime s_{jk}^{\prime} \Sigma_{jk}^{\prime}}^{(jk-)} 
e^{i\bd{P}^\prime \cdot \bd{R}^\prime }|V_{0i}|\Psi^{(JM)}
e^{i\bd{P} \cdot \bd{R}} \rangle \; ,
\label{trans}
\end{equation}
where $J$ and $M$ are the total angular momentum of the projectile and
its projection on the $z$--axis, $s_{jk}^{\prime}$ and
$\Sigma_{jk}^{\prime}$ are the spin and its third component of the
two--body system formed by particles $j$ and $k$ after the collision,
$s_i^{\prime}$ and $\Sigma_i^{\prime}$ are the spin and projection
quantum numbers of particle $i$ (for convenience we assume a spin zero
target), and $\phi _{\bd{p}_{0i}^{\prime }\Sigma_i^{\prime}}^{(0i-)}$
and $\phi _{\bd{p}_{jk}^{\prime }s_{jk}^{\prime}
\Sigma_{jk}^{\prime}}^{(jk-)}$ are the distorted wave functions of 
the two final two--body systems.  We use plane waves $e^{i\bd{P}^\prime
\cdot \bd{R}^\prime }$ and $e^{i\bd{P} \cdot \bd{R} }$ because these 
relative momenta are
much higher and the plane wave approximation is suitable.

In the frame of the projectile ($\bd{p}_i+\bd{p}_j+\bd{p}_k=0$), and 
assuming that the participant $i$ has spin 0 or 1/2, the previous
transition amplitude leads to the following differential 
diffraction cross section for the process in Fig.~\ref{fig1}
\begin{equation}
 \frac{d^9\sigma _{el}^{(i)}(\bd{P}^{\prime },
\bd{p}_{jk}^{\prime },\bd{q})}
{ d\bd{P}^{\prime } d\bd{p}_{jk}^{\prime } d\bd{q} }
  =    \frac{d^3\sigma _{el}^{(0i)}(\bd{p}_{0i}
  \rightarrow  \bd{p}_{0i}^{\prime})}
 {d\bd{q}} \;
\frac{ P_{dis}(\bd{q})}{2 J+1} \sum_{M s_{jk}^{\prime}\Sigma_{jk}^{\prime}
\Sigma_i^{\prime} }
 |M_{s_{jk} \Sigma_{jk}^{\prime} \Sigma_i^{\prime}}^{(JM)}|^{2} \; ,
\label{elas}
\end{equation}
\begin{equation}
 M_{s_{jk}^{\prime} \Sigma_{jk}^{\prime} \Sigma_i^{\prime}}^{(JM)} = \langle
\phi _{\bd{p}_{jk}^{\prime }s_{jk}^{\prime} \Sigma_{jk}^{\prime}}^{(jk-)} e^{i
\bd{p}_{i,jk}\bd{r}_{i,jk}}\chi _{s_{i}^{\prime}\Sigma _{i}^{\prime}}
|\Psi^{(JM)}\rangle \; .
\label{overl}
\end{equation}
Note that the final two--body wave function $\phi_{ \bd{p}_{0i}^\prime  
\Sigma^\prime_i}^{(0i-)}$ is contained in the differential elastic
cross section for the participant--target scattering
$d^3\sigma^{(0i)}_{el}/d\bd{q}$. The two--body wave function
$\phi _{\bd{p}_{jk}^{\prime }s_{jk}^{\prime} \Sigma_{jk}^{\prime}}^{(jk-)}$ 
is computed as in \cite{gar97b} and the appropriate phase factor
is included in the computation of the overlap matrix element.
When the participant $i$ in the sense of the optical model is absorbed
by the target the corresponding differential absorption cross section
factorizes as
\begin{equation} 
\label{abs}
 \frac{d^6\sigma _{abs}^{(i)}(\bd{P}^{\prime },\bd{p}_{jk}^{\prime })}
{ d\bd{P}^{\prime } d\bd{p}_{jk}^{\prime } }
=   \sigma _{abs}^{(0i)}(p_{0i}) \;
\frac{1}{2 J+1} \sum_{M s_{jk}^{\prime}\Sigma_{jk}^{\prime}\Sigma_i^{\prime} }
 |M_{s_{jk} \Sigma_{jk}^{\prime} \Sigma_i^{\prime}}^{(JM)}|^{2} \; .
\end{equation}

The final state wave function used in eq.(\ref{trans}) is not
orthogonal to the initial wave function.  This means that our final
state contains a non-zero component of three--body bound state plus
target, that represents elastic scattering of the halo nucleus as a
whole, and has to be removed. This is done in eq.(\ref{elas}) by means
of the function $P_{dis}(\bd{q})$, that is the probability of
dissociation of the three--body halo system. For elastic scattering
the final state wave function of the halo nucleus is $e^{i \bd{q}_{cm}
\cdot \bd{r}_{i,jk}} \Psi^{(JM)}$, where $\bd{q}_{cm}= \bd{q}
(m_j+m_k)/(m_i+m_j+m_k)$ is the momentum transfer into the
participant--spectators relative motion described by the coordinate
$\bd{r}_{i,jk}$. Therefore the probability of elastic scattering of
the three--body projectile as a whole is given by the square of the
overlap between the initial and final three--body wave functions. This
implies that the probability of dissociation is given by \cite{han87}
\begin{equation}
P_{dis}(\bd{q})=1-|\langle \Psi^{(JM)}|e^{i \bd{q}_{cm} \cdot \bd{r}_{i,jk}}
                                      |\Psi^{(JM)}\rangle|^2 \; ,
\label{probdis}
\end{equation}
which vanishes as $q^2 (\propto \bd{q}_{cm}^2)$ for small values of
the momentum transfer.

In eqs.(\ref{elas}) and (\ref{abs}) $d^3\sigma _{el}^{(0i)} / d\bd{q}$
and $\sigma _{abs}^{(0i)}$ are the differential elastic and absorption
cross sections for the participant--target scattering, respectively
\cite{sit71}. We use a central
participant--target nuclear potential $V_N(r)$, and assuming that the
participant and the target have charges $Z_i$ and $Z_0$, respectively,
the participant--target interaction is
\begin{equation}
V_{0i}(r)=\frac{Z_0 Z_i \alpha}{r} + V_N(r) \; ,
\label{v0i}
\end{equation} 
where $\alpha=e^2/\hbar c$ is the electromagnetic coupling constant. 

The cross sections arising from the interaction $V_{0i}$ in
eq.(\ref{v0i}) are obtained after integration of eqs.(\ref{abs}) and
(\ref{elas}) when particle $i$ is absorbed and elastically scattered,
respectively.

\subsection{Three by one separation in the final state}

The scheme shown in Fig.~\ref{fig1} arises when the halo breakup is caused
by a sufficiently violent and fast participant-target reaction. However this
is not true in processes in which contributions from small momentum
transfers are dominating. In these cases the small kick of the participant
excites the halo rather gently and the breakup process proceeds via the
created wave packet. An obvious example is halo breakup on a heavy target
caused by the Coulomb participant--target interaction, which clearly is a
dominating process, especially at low beam energies \cite{gar00a}.

\begin{figure}[tbp]
\centerline{\psfig{figure=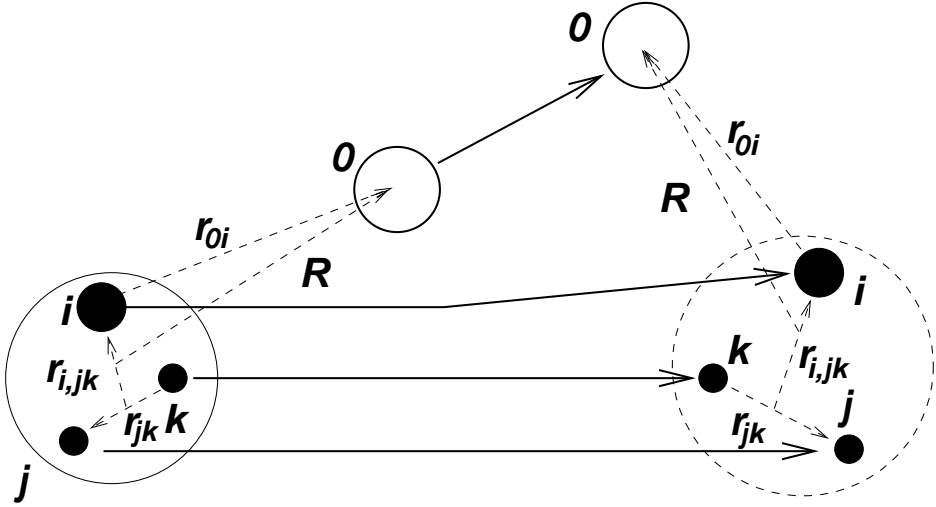,width=14cm,%
bbllx=-1.0cm,bblly=-6.cm,bburx=15.4cm,bbury=8cm,angle=0}}
\vspace*{-5cm}
\caption{ Sketch of the reaction and coordinates used for the final state
three by one separation. The notation is as in Fig.~\ref{fig1}.}
\label{fig1a}
\end{figure}

This type of gentle breakup reaction mechanism requires that the final state
is that component of the three-body continuum halo wave function which at
large distances approaches plane waves with relative momenta specified by $
\bd{p}_{jk}^{\prime }$ and $\bd{p}_{i,jk}^{\prime }$. The corresponding
reaction is sketched in Fig.~\ref{fig1a}. The transition matrix in eq.(\ref
{trans}) should then be replaced by 
\begin{equation}
T^{(i)}=\langle\Psi_{{\bf p}_{jk}^{\prime }{\bf p}_{i,jk}^{\prime }}^{(-)}
\;\chi _{s_{i}^{\prime}\Sigma _{i}^{\prime}} \;\chi _{s_{jk}^{\prime}
\Sigma_{jk}^{\prime}}\; e^{i\bd{P}^\prime\cdot\bd{R}} |V_{0i}(r_{0i})| \Psi_{
{\bf P}}^{(+)}(R,{\bf r}_{jk},{\bf r}_{i,jk})\rangle  \label{trans2}
\end{equation}
where $\bd{R}$ now is the relative coordinate between the three--body system
and the target, and $\bd{P}$ and $\bd{P}^\prime$ are its conjugate momenta
respectively in the initial and final state -- note that in section 2.1 $
\bd{P}^\prime$ is the conjugate momentum of the relative radial coordinate
between the center of mass of the two final two--body systems (see Fig.~\ref
{fig1}), while now $\bd{P}^\prime$ is the conjugate momentum of the relative
radial coordinate between the final three--body system and the target. The
two spin wave functions $\chi$ are labeled by their respective final state
total and projection quantum numbers $(s_{i}^{\prime}\Sigma_{i}^{\prime})$
and $(s_{jk}^{\prime} \Sigma_{jk}^{\prime})$ defined earlier.

The distorted wave $\Psi _{{\bf P}}^{(+)}$ satisfies the equation 
\begin{equation}
\left[ -\frac{\hbar ^{2}}{2M}\frac{\partial ^{2}}{\partial\bf{R}^{2}}
+V_{0i}(r_{0i})+H({\bf r}_{jk},{\bf r}_{i,jk})-E\right] \Psi _{{\bf P}
}^{(+)}=0,  \label{dscat}
\end{equation}
where $M$ is the projectile-target reduced mass and $H({\bf r}_{jk},{\bf
r}_{i,jk})$ is the intrinsic hamiltonian of the projectile. A gentle kick
by a long range Coulomb potential implies that the halo absorbs the transfered
momentum as a whole, meaning that the adiabatic approximation should be
adequate for this reaction type.

We therefore as in \cite{tos98} replace the intrinsic hamiltonian $H({\bf r
}_{jk},{\bf r}_{i,jk})$ with a constant, namely the ground state energy $-B$
of the halo.  This ground state energy, typically a fraction of an MeV, is
much smaller than the total energy $E$, typically of the order of hundred
MeV, and can therefore be neglected. The internal
coordinates ${\bf r}_{jk}$, ${\bf r}_{i,jk}$ are now no longer dynamical
variables in the equation but parameters and therefore $d{\bf R}=d{\bf r}
_{0i}$ leading to
\begin{equation}
\left[ -\frac{\hbar ^{2}}{2M}\frac{\partial ^{2}}{\partial {\bf r}_{0i}^{2}}
+V_{0i}(r_{0i})-E\right] \Psi _{{\bf P}}^{(+)}=0.
\end{equation}
The solution to this equation  is 
\begin{equation}
\Psi _{{\bf P}}^{(+)}=\psi _{{\bf P}}^{(+)}(\,{\bf r}_{0i})\psi ({\bf r}
_{jk},{\bf r}_{i,jk}),
\end{equation}
where $\psi _{{\bf P}}^{(+)}(\,{\bf r}_{0i})$ is the distorted wave
describing scattering of a particle with mass $M$ in the potential $V_{0i}$
and the function $\psi ({\bf r}_{jk},{\bf r}_{i,jk})$ is determined from the
large ${\bf r}_{0i}$ asymptotics 
\begin{equation}
\Psi _{{\bf P}}^{(+)}\stackrel{r_{0i}\rightarrow \infty }{\longrightarrow }
\left( e^{i{\bf Pr}_{0i}}+f(\theta )\frac{e^{iPr_{0i}}}{r_{0i}}\right) \psi (
{\bf r}_{jk},{\bf r}_{i,jk}).
\end{equation}
Indeed, the plane-wave term of this expression describes a projectile in the
ground state $\Psi ^{(JM)}$ moving with the momentum $P$ against the target,
i.e. 
\begin{equation}
e^{i{\bf Pr}_{0i}}\psi ({\bf r}_{jk},{\bf r}_{i,jk})=
e^{i{\bf P}\cdot{\bf R}}\Psi ^{(JM)},
\end{equation}
which gives 
\begin{equation}
\psi ({\bf r}_{jk},{\bf r}_{i,jk})=
e^{i{\bf P}\cdot{\bf R-}i{\bf P}\cdot{\bf r}_{0i}}\Psi
^{(JM)}=e^{i\mu _{i}{\bf P}\cdot{\bf r}_{i,jk}}\Psi ^{(JM)},
\end{equation}
where $\mu _{i}=(m_{j}+m_{k})/(m_{i}+m_{j}+m_{k})$. The matrix element $
T^{(i)}$ in eq. (\ref{trans2}) then becomes 
\begin{eqnarray}
T^{(i)} &=&\langle \Psi _{{\bf p}_{jk}^{\prime }{\bf p}_{i,jk}^{\prime
}}^{(-)}\;\chi _{s_{i}^{\prime }\Sigma _{i}^{\prime }}\;\chi
_{s_{jk}^{\prime }\Sigma _{jk}^{\prime }}\;e^{i\mu _{i}{\bf P}^{\prime }
\cdot{\bf 
r}_{i,jk}}e^{i{\bf P}^{\prime }\cdot{\bf r}_{0i}}|  \nonumber \\
&\times &V_{0i}(r_{0i})|\psi _{{\bf P}}^{(+)}(\,{\bf r}_{0i})e^{i\mu _{i}
{\bf P}\cdot{\bf r}_{i,jk}}\Psi ^{(JM)}\rangle ,
\end{eqnarray}
which factorizes as 
\begin{eqnarray}
T^{(i)} &=&\langle \Psi _{{\bf p}_{jk}^{\prime }{\bf p}_{i,jk}^{\prime
}}^{(-)}\;\chi _{s_{i}^{\prime }\Sigma _{i}^{\prime }}\;\chi
_{s_{jk}^{\prime }\Sigma _{jk}^{\prime }}\;|e^{i{\bf q}_{cm}\cdot{\bf r}
_{i,jk}}|\Psi ^{(JM)}\rangle   \nonumber \\
&\times &\langle e^{i{\bf P}^{\prime }\cdot{\bf r}_{0i}}\left|
V_{0i}(r_{0i})\right| \psi _{{\bf P}}^{(+)}(\,{\bf r}_{0i})\rangle ,
\end{eqnarray}
where ${\bf q}_{cm}=\mu _{i}{\bf q}$ and ${\bf q}={\bf P}^{\prime
}-{\bf P}$. The second factor is the matrix element $T_{el}^{(0i)}$
of the elastic scattering on the potential $V_{0i}$ of a particle with
mass $M$ -- that is the projectile-target scattering with the potential
$V_{0i}$.

In addition to the breakup amplitude we obtain the elastic scattering of the
halo as a whole simply by substituting the ground state wave function $\Psi
^{(JM)}$ instead of the continuum wave function in the transition matrix
element in eq.(\ref{trans2}). This then leads to 
\begin{equation}
T_{el}^{(i)}=T_{el}^{(0i)}\;\langle \Psi ^{(JM)}|e^{i\bd{q}_{cm}\cdot \bd{r}
_{i,jk}}|\Psi ^{(JM)}\rangle \;.  \label{elast}
\end{equation}

For a two-neutron halo the largest contribution to the three-by-one
final state is given at large impact parameters by the long-range Coulomb
interaction between the charged core and the target. Considering only this
main contribution we can write the breakup cross-section (with $i$ designating
the core) as
\begin{eqnarray}
&&\frac{d^{9}\sigma }{d\bd{P}^{\prime }d\bd{p}_{jk}^{\prime }d\bd{q}}=\frac{
d^{3}\sigma _{Ruth}^{(0p)}}{d\bd{q}}\;  \nonumber \\
&&\times \frac{1}{2J+1}\sum_{M,\Sigma _{i}^{\prime },s_{jk}^{\prime },\Sigma
_{jk}^{\prime }}|\langle \Psi _{\bd{p}_{jk}^{\prime },\bd{p}_{i,jk}^{\prime
}}^{(-)}\;\chi _{s_{i}^{\prime }\Sigma _{i}^{\prime }}\;\chi
_{s_{jk}^{\prime }\Sigma _{jk}^{\prime }}\;|e^{i\bd{q}_{cm}\cdot \bd{r}
_{i,jk}}|\Psi ^{(JM)}\rangle |^{2}\;,  \label{elascoul}
\end{eqnarray}
where the spin of the core is 0 or 1/2 and $d^{3}\sigma _{Ruth}^{(0p)}$  is
simply the Rutherford projectile-target cross section.

When the two--body interactions between halo fragments in the final state
are neglected in eq.(\ref{elascoul}) the matrix element simply reduces to
the Fourier transform of the three-body bound state wave function with the
momentum $\bd{p}_{i,jk}$ shifted by $\bd{q}_{cm}$.

\subsubsection{The continuum three-body wave function}

The three-body continuum wave function $\Psi_{\bd{p}_{jk}^{\prime
},\bd{p}_{i,jk}^{\prime }}^{(-)}$ in eq.(\ref{elascoul}) has to be
computed. We first introduce the Jacobi coordinates (see
e.g. \cite{zhu93}) ${\bf x}$ and ${\bf y}$ and their conjugate momenta
${\bf k}_x$ and ${\bf k}_y$ such that ${\bf k}_x \cdot {\bf x} \equiv
\bd{r}_{jk}^{\prime } \cdot \bd{p}_{jk}^{\prime }$ and ${\bf k}_y
\cdot {\bf y} \equiv \bd{p}_{i,jk}^{\prime } \cdot
\bd{r}_{i,jk}^{\prime }$. The corresponding hyperspherical coordinates
$\rho, \Omega$ are then also available. We omit here the label $i$
specifying which set of Jacobi coordinates.  We then define the
continuum three-body wave function $\Psi _{{\bf k}_{x}{\bf
k}_{y}}^{(-)}({\bf x},{\bf y})$ with the asymptotic behavior of a plane wave
in six dimensions and an incoming hyperspherical wave
\begin{equation}
\Psi_{{\bf k}_{x}{\bf k}_{y}}^{(-)}({\bf x},{\bf y})\rightarrow 
\exp (i{\bf k}_{x} \cdot {\bf x}+i{\bf k}_{y} \cdot {\bf y}) + 
f^{*}(\Omega _{\kappa },\Omega )\frac{
\exp (-i\kappa \rho )}{\rho ^{5/2}},  \label{a0}
\end{equation}
where we have introduced the 6-momentum $ {\bf \kappa } \equiv \{{\bf
k}_{x},{\bf k} _{y}\}$ and the corresponding hyperangular variables
$\kappa =\sqrt{ k_{x}^{2}+k_{y}^{2}}$ and $\Omega _{\kappa }$.  The
total energy is then given by $\kappa$ as $E=\hbar ^{2}\kappa
^{2}/(2m)$.

We shall now express the function $\Psi _{{\bf k}_{x}{\bf k}_{y}}$ in
terms of our adiabatic hyperspherical basis functions $\Phi
_{JMn}(\rho ,\Omega )$, see e.g.\cite{nie00,gar96,gar97}.  The basis
functions $\Phi _{JMn}$ are labeled according to their asymptotic
behavior, which means approaching the corresponding hyperspheric
harmonics, i.e. $\Phi _{JMn}(\rho ,\Omega )\rightarrow {\bf Y
}_{Q}(\Omega)$ for large $\rho$. Here we use the notation ${\bf Y
}_{Q}(\Omega)$ for a hyperspheric harmonic coupled with the
appropriate spin functions to the total angular momentum, where $Q
\equiv \{K,l_{x},l_{y},L,s_x,s_y,S,J,M\}$ denotes the full set of
hyperradial quantum numbers, where $s_x = s_{jk}$ and $s_y = s_{i}$.

We now first shift the dependence of the function upon the five
hyperangles $\Omega _{\kappa }$ of the momentum $\kappa$ into the
basis $Y_{\xi}(\Omega _{\kappa })$ ($\xi \equiv\{K,l_x,l_y,L\}$) of the
ordinary non-spin coupled hyperspheric harmonic.  Thus we get
\begin{equation}
\Psi_{{\bf k}_{x}{\bf k}_{y}}^{(-)}({\bf x},{\bf y})
 \; \chi _{s_{i}^{\prime}\Sigma _{i}^{\prime}} 
 \; \chi _{s_{jk}^{\prime} \Sigma_{jk}^{\prime}} \; = 
\left[\sum_{\xi}\psi_{\xi}
({\bf x},{\bf y}) {Y }_{\xi}^{*}(\Omega _{\kappa }) \right]
 \chi _{s_{i}^{\prime}\Sigma _{i}^{\prime}} 
 \; \chi _{s_{jk}^{\prime} \Sigma_{jk}^{\prime}} \; ,
\end{equation}
where the function $\psi_{\xi}({\bf x},{\bf y})$ now can be expanded
in terms of our complete set of angular solutions $\Phi _{JMn}(\rho
,\Omega )$ to the Faddeev equations i.e.
\begin{equation}
\psi_{\xi}({\bf x},{\bf y})  \chi _{s_{i}^{\prime}\Sigma _{i}^{\prime}} 
 \; \chi _{s_{jk}^{\prime} \Sigma_{jk}^{\prime}} = \rho ^{-5/2}
\sum_{JMn}F_{JMn}^{(q)}(\rho )\Phi _{JMn}(\rho ,\Omega ) \; ,  \label{a2}
\end{equation}
where $ q \equiv \{\xi,s_{jk}^{\prime},\Sigma_{jk}^{\prime},
s_{i}^{\prime},\Sigma _{i}^{\prime}\}$ is the set of quantum numbers
equivalent to $Q$ in the decoupled spin basis.

Inserting expansion eq.(\ref{a2}) into the Schr\"{o}dinger equation we
obtain, after multiplication by $\langle \Phi _{JMn}(\rho ,\Omega )|$
and subsequent angular integration, our usual hyperradial equations
for $F_{JMn}^{(q)}(\rho )$ \cite{nie00,gar96,gar97}
\begin{eqnarray}
&\left( -\frac{\partial ^{2}}{\partial \rho ^{2}}+\frac{\lambda _{n}(\rho )+
\frac{15}{4}}{\rho ^{2}}+Q_{nn}-\frac{\hbar ^{2}\kappa ^{2}}{2m}\right)
F_{JMn}^{(q)}(\rho ) \nonumber \\
& = \sum_{n^{\prime }\neq n}\left( -2P_{nn^{\prime }}
\frac{\partial }{\partial \rho }-Q_{nn^{\prime }}\right) 
F_{JMn^{\prime }}^{(q)}(\rho ).
\end{eqnarray}

The asymptotic form eq.(\ref{a0}) can be expanded in terms of the
hyperspheric harmonics ${Y }_{\xi}(\Omega _{\kappa })$ using
\begin{equation}
\exp(i{\bf k}_x{\bf x}+i{\bf k}_y{\bf y}) = \frac{(2\pi)^3}{(\kappa\rho)^2}
\sum_{\xi} i^K J_{K+2}(\kappa\rho) 
{Y }_{\xi}^{*}(\Omega _{\kappa })  { Y }_{\xi}(\Omega) \; ,
\end{equation}
which implies the asymptotic behavior 
\begin{eqnarray}
 \Psi_{{\bf k}_{x}{\bf k}_{y}}^{(-)}({\bf x},{\bf y})\rightarrow 
\sum_{\xi \xi^{\prime }}\left( \delta _{\xi \xi^{\prime }}
\frac{(2\pi )^{3} \; i^{K}}{
(\kappa \rho )^{2}}J_{K+2}(\kappa \rho ) \right. \nonumber \\ \left.
 + f_{\xi \xi^{\prime }}^{*}\frac{
\exp (-i\kappa \rho )}{\rho ^{5/2}}\right) Y_{\xi^{\prime}} (\Omega )
 {Y }_{\xi}^{*}(\Omega _{\kappa }) \; ,  \label{a1}
\end{eqnarray}
Then the asymptotic boundary condition for the hyperradial function
$F_{JMn}^{(q)}(\rho )$ is found from eq.(\ref{a1})
\begin{eqnarray}
\rho ^{-5/2}F_{JMn}^{(q^{\prime})}(\rho )\rightarrow \sum_{\xi}
\left[\frac{(2\pi
)^{3} \; i^{K}}{(\kappa \rho )^{2}}J_{K+2}(\kappa \rho )
\delta _{\xi \xi^{\prime}}+f_{\xi \xi^{\prime}}^{*}
\frac{\exp (-i\kappa \rho )}{\rho ^{5/2}}\right] \nonumber \\
 \times \langle {\bf Y}_{Q} | Y_{\xi}  
\chi _{s_{i}^{\prime}\Sigma _{i}^{\prime}} 
 \; \chi _{s_{jk}^{\prime} \Sigma_{jk}^{\prime}}\rangle \; .
\end{eqnarray}
This boundary condition is equivalent to the scattering boundary
condition used in \cite{cob98}.

The transition matrix element in eq.(\ref{elascoul}) reduces when
$\bd{q}_{cm} \cdot \bd{r}_{i,jk} \ll 1 $ to 
\begin{equation}
 i \bd{q}_{cm} \cdot 
\langle \Psi_{\bd{p}_{jk}^{\prime},\bd{p}_{i,jk}^{\prime }}
 \; \chi _{s_{i}^{\prime}\Sigma _{i}^{\prime}} 
 \; \chi _{s_{jk}^{\prime} \Sigma_{jk}^{\prime}} \;
| \bd{r}_{i,jk} |   \Psi^{(JM)}\rangle \; ,
\end{equation}
which after squaring and summation over magnetic quantum numbers gives
the usual factorized expression in terms of the $B(E1)$ dipole
strength function and the cross section \cite{cob98}. Before expansion
this matrix element also included all the higher multipole
transitions. They could in principle arise from the nuclear as well as
the Coulomb interaction.

\subsection{Momentum cutoff}
\label{sec23}

When the participant is charged the elastic cross section
eq.(\ref{elascoul}) for $q \rightarrow 0$ is dominated by the Coulomb
cross section, which after integration over the
directions of $\bd{q}$ diverges like $q^{-3}$.  Since the square of
the matrix element in eq.(\ref{elascoul}) vanishes as $q^2$ for small
$q$, the integration of eq.(\ref{elascoul}) over $q$ leads to a
logarithmic divergence of this elastic cross section.  However, to
produce dissociation, the energy transferred from target to
participant ($\delta E \equiv
\sqrt{\bd{p}_0^2+m_0^2}-\sqrt{\bd{p}_0^{\prime 2}+m_0^2}$) must be
larger than the three--body binding energy $B$.  When $\bd{p}_0$ and
$\bd{q}=\bd{p}_0-\bd{p}^\prime_0$ are parallel $\delta E$ is
maximized.  For small $B$ compared to the target rest mass $\delta
E=B$ implies that $qc \approx q_L c \equiv B \sqrt{1+m_0^2c^2/p_0^2}$,
which reduces to $B/v$ in the non--relativistic limit.  Thus $q$ must
be larger than $q_L$ to produce dissociation.

Another formal divergence arises for the Coulomb interaction at large
impact parameters (low momentum transfer), where adiabatic motion only
allows virtual excitations excluding dissociation.  If the collision
time $\Delta t$ is long compared to the orbital period of motion $T$
inside the projectile no transfer of energy occurs. These processes
are excluded by the adiabatic cutoff.  
The decisive parameter is \cite{ber88}
\begin{equation}
 \xi_a \equiv \frac{ \hbar \omega_a b}{\gamma \beta \hbar c} \; ,
\label{times}
\end{equation}
where $\beta=v/c$, $\gamma=(1-\beta^2)^{-1/2}$ and $\hbar \omega_a $ is
the adiabatic cutoff energy, i.e. the maximum value of the equivalent
photon energy for a given impact parameter $b$ estimated in a sharp cutoff
model \cite{ber88}. Then $\hbar \omega_a \approx B$, where $B$ is the
three-body binding energy, should be a reasonable estimate as the lowest
energy needed for breakup. However, the softness of the electromagnetic
excitation modes could influence this sharp cutoff value, but the effect
on the cross sections is still only logarithmic. We leave the precise
choice of $\hbar \omega_a $ for adjustments in the numerical computations.

In previous publications \cite{gar00a,gar00b} we used $\hbar \omega_a = B_{ps} /
\pi$, where $B_{ps}$ is the binding energy between participant and the
system consisting of the two spectators, i.e., $B_{ps} = B - B_{2s}$,
where $B_{2s}$ is the two--body binding energy of the two spectators. In a
Borromean system $B_{2s}$ is negative. This is a misleading connection for
a slow reaction where the full projectile is adiabatically excited.

From \cite{jac75} the following classical
relation between the momentum transfer $q$ and the impact parameter
$b$ is derived:
\begin{equation}
q=\frac{Z_i Z_0 e^2}{b} \frac{p}{E_{kin}}  \; ,
\label{adiab}
\end{equation}
where $p$ is the momentum of the projectile and $E_{kin}$ its kinetic energy.

The adiabatic cutoff condition $\xi_a=1$ and use of
eqs.(\ref{times}) and (\ref{adiab}) leads then to $q>q_a$, where
\begin{equation}
q_a = \frac{Z_0Z_i e^2}{(\hbar c)^2} 
              \frac{\hbar \omega_a}{\gamma -1} \; .
\label{coulcut}
\end{equation}
Therefore integration of eq.(\ref{elascoul}) has to include only
values of the momentum transfer $q$ larger than the largest of $q_L$
and $q_a$. The divergence then disappears.

\section{Model and method for spatially extended particles}
\label{sec3}

The two different final state divisions described in Section
\ref{sec2} are introduced for physics reasons and not for convenience.
The underlying idea is that dominating reaction contributions arise
from (optical model type of) absorption, which implies that the final
state consists of the spectators and the participant-target system,
and elastic core-target Coulomb scattering at low momentum transfer
leading to a final state containing all three halo particles.  This
division is dictated by the reaction mechanism and seems to be
unavoidable in a few-body treatment where the core degrees of freedom
are neglected.  Furthermore, the finite extension of the projectile
constituents and the target also requires a similar division according
to the number of projectile constituents simultaneously colliding with
the target.

\subsection{Reaction scenarios}

\begin{figure}
\centerline{\psfig{figure=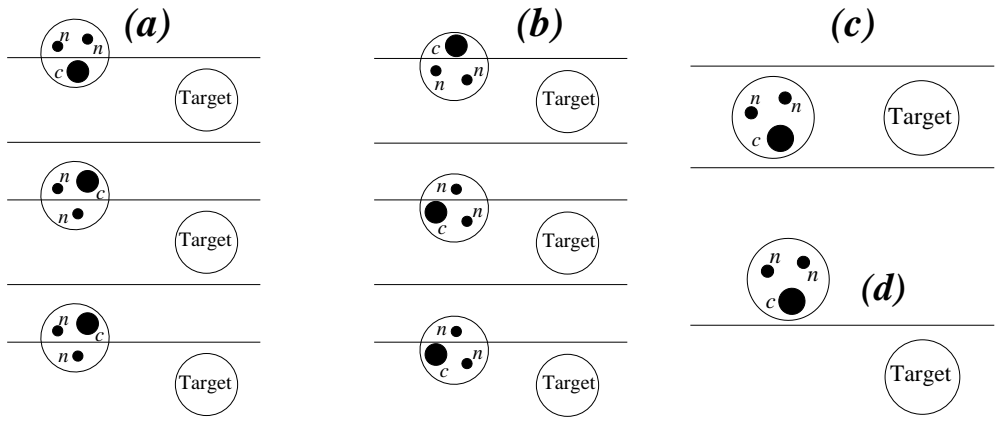,width=7.0cm,%
bbllx=7.4cm,bblly=-10.5cm,bburx=23.8cm,bbury=13.5cm,angle=0}}
\vspace*{-4.5cm}
\caption[]{ Scheme of the possible scenarios for the collision between
the three--body projectile and the target. The notation indicates a
halo consisting of two neutrons and a core.}
\label{fig2}
\end{figure}

Let us extend the applicability of the inert particle model and
incorporate effects of the core degrees of freedom and the finite
sizes of the constituents. First we assume short--range interactions 
between each of the halo constituents and the target. Then we start by
describing the individual constituent--target interaction in the black 
disk model, according to which the target absorbs the constituent inside 
a cylinder with the axis along the beam direction and otherwise leaves
untouched.  The radius of this cylinder is related to the range of the
interaction and approximately equal to the sum of target and
constituent radii. For the collision between the halo nucleus and the
target we have then the four possibilities shown in Fig.~\ref{fig2}:

\begin{itemize}
\item[(a)] Only one of the halo constituents, the participant, is
inside its cylinder. The other two constituents are mere spectators,
and survive untouched after the collision. The three constituents of
the halo projectile give rise to three contributions of this kind.
\vspace*{3mm}

\item[(b)] Two of the halo constituents are participants. The third one
is spectator. Again we have three possible reactions of this kind.
\vspace*{3mm}

\item[(c)] All the three halo constituents are inside their corresponding
cylinders.
\vspace*{3mm}

\item[(d)] All the three halo constituents are outside their
cylinders. For short--range interactions this means that the whole
projectile is untouched by the target, and this process would only
contribute to the elastic cross section.
\end{itemize}

The scheme in Fig.~\ref{fig2} is only directly valid for short-range
interactions and therefore not applicable for charged halos. This is a
severe deficiency as even a two--neutron halo nucleus colliding with a
heavy target involves a core-target Coulomb interaction that by far
can dominate the cross section.

To incorporate long--range interactions into the model we do the
following: In the reactions shown in Figs.~\ref{fig2}a, \ref{fig2}b, and
\ref{fig2}c the participant--target interaction contains two parts,
the full short--range interaction and the long--range potential
related to small impact parameters (large momentum transfer). The
large impact parameter part of the long--range potentials (small
momentum transfer) corresponds to reactions as the one shown in
Fig.~\ref{fig2}d.

In the particular case of two--neutron halo fragmentation this means
that the momentum transfer between target and core is divided into
large and small values corresponding to short and long impact
parameters, respectively. Large momentum transfer contributions to the
cross sections correspond to the processes shown in Figs.~\ref{fig2}a,
\ref{fig2}b, and \ref{fig2}c. They contain the nuclear core-target
interaction and the large momentum transfer part of the Coulomb
core--target interaction.  They are described in Section 2.1, and the
integral of eq.(\ref{elas}) over the momentum transfer involves values
of $q$ larger than a certain value $q_g$ ($q_g < q <\infty$).  Low
momentum transfer contributions correspond to the process shown in
Fig.~\ref{fig2}d, where none of the halo constituents are inside their
cylinders, and only the core-target Coulomb interaction contributes.
Cross sections are obtained after integration of eq.(\ref{elascoul}),
where the momentum transfer varies between
$q_{min}=\mbox{max}\{q_L,q_a\}$ and $q_g$.

Low and large impact parameters are divided by the value $b_g=
R_0+R_c+ \pi a /2$, see \cite{ber88}, where $R_0$ and $R_c$ are charge
root mean square radii of the target and the core, and $a=\hbar c Z_0
Z_c e^2 /2 E_{kin}$ is half the distance of closest core-target
approach.  Use of eq.(\ref{adiab}) and the fact that
$p/E_{kin}=(\gamma+1)/\gamma \beta c$ then leads to the value $q_g$ of
the momentum separating low ($q<q_g$, $b>b_g$) and large momentum
transfer ($q>q_g$, $b<b_g$), i.e.
\begin{equation}
q_g = \frac{\hbar Z_0Z_ce^2 }{b_g}
           \frac{\gamma +1}{\gamma \beta} \; .
\label{qg}
\end{equation}

\subsection{Cross sections}
\label{sub30}

According to the number of particles surviving in the final state we
have the possible cross sections: $\sigma_0$, $\sigma_{n}$,
$\sigma_c$, $\sigma_{nn}$, $\sigma_{nc}$ and $\sigma_{nnc}$, where
$\sigma_0$ means no particles in the final state, and the indexes $c$
and $n$ refer to the core and the neutrons surviving in the final
state. If some of the particles are absent in the final state they
have been absorbed by the target in the optical model sense.

The probabilities for finding the core and one halo neutron inside
their cylinders are denoted $P_c$ and $P_n$, respectively.  These
probabilities can be split in two parts, the probability of being
absorbed by the target and the probability of being elastically
scattered by the target: $P_c=P_c^{ab}+P_c^{el}$ and
$P_n=P_n^{ab}+P_n^{el}$.

\begin{table}
\begin{center}
\caption{Contributions to the different cross sections from the
reactions shown in Figs.~\ref{fig2}a, \ref{fig2}b and \ref{fig2}c,
where the nuclear interaction between the halo constituents plus the
large momentum transfer (low impact parameter) part of the Coulomb
core--target interactions are included. Two crosses mean that the
contribution must be included twice due to the possible exchange of
the two halo neutrons. The cross section $\sigma_{nnc}$ contains also
the contribution from the low momentum transfer (large impact
parameter) part of the Coulomb core--target interaction sketched in
Fig.~\ref{fig2}d.}
\label{table1}
\vspace*{0.3cm}
\begin{tabular}{|c|c|cccccc|}
\hline
Reaction & Probability &
$\sigma_0$ & $\sigma_n$ & $\sigma_c$ & $\sigma_{nn}$ & 
              $\sigma_{nc}$ & $\sigma_{nnc}$  \\ \hline
  & $P_c^{el}(1-P_n)(1-P_n)$   &   &   &   &   &   & X \\
  & $P_c^{ab}(1-P_n)(1-P_n)$   &   &   &   & X &   &   \\
Fig.\ref{fig2}a  
  & $P_n^{el}(1-P_n)(1-P_c)$   &   &   &   &   &   & XX\\
  & $P_n^{ab}(1-P_n)(1-P_c)$   &   &   &   &   & XX&   \\ \hline
  & $P_c^{el}P_n^{el}(1-P_n)$  &   &   &   &   &   & XX\\
  & $P_c^{el}P_n^{ab}(1-P_n)$  &   &   &   &   & XX&   \\
  & $P_c^{ab}P_n^{el}(1-P_n)$  &   &   &   & XX&   &   \\
Fig.\ref{fig2}b
  & $P_c^{ab}P_n^{ab}(1-P_n)$  &   & XX&   &   &   &   \\
  & $P_n^{el}P_n^{el}(1-P_c)$  &   &   &   &   &   & X \\
  & $P_n^{el}P_n^{ab}(1-P_c)$  &   &   &   &   & XX&   \\
  & $P_n^{ab}P_n^{ab}(1-P_c)$  &   &   & X &   &   &   \\ \hline
  & $P_c^{el}P_n^{el}P_n^{el}$ &   &   &   &   &   & X \\
  & $P_c^{el}P_n^{el}P_n^{ab}$ &   &   &   &   & XX&   \\
  & $P_c^{el}P_n^{ab}P_n^{ab}$ &   &   & X &   &   &   \\
Fig.\ref{fig2}c
  & $P_c^{ab}P_n^{el}P_n^{el}$ &   &   &   & X &   &  \\
  & $P_c^{ab}P_n^{el}P_n^{ab}$ &   & XX&   &   &   &  \\
  & $P_c^{ab}P_n^{ab}P_n^{ab}$ & X &   &   &   &   &  \\ \hline
\end{tabular}
\end{center}
\end{table}

When the three--body halo projectile collides with the target, each of
the reactions shown in Fig.~\ref{fig2}a occur with the probability
$P_i(1-P_j)(1-P_k)$, where the index $i$ refers to the participant.
In the same way each of the reactions shown in Fig.~\ref{fig2}b and
Fig.~\ref{fig2}c occur with the probabilities $P_iP_j(1-P_k)$ and
$P_iP_jP_k$.  In this way, the total probability of a given process is
obtained by adding the probabilities of all the processes in
Fig.~\ref{fig2} producing precisely the corresponding group of halo
constituents in the final state.

For instance, the probability of a process in which only the core
survives in the final state (processes giving rise to $\sigma_c$)
contains two terms, $P_n^{ab}P_n^{ab}(1-P_c)$, that comes from one of
the reactions shown in Fig.~\ref{fig2}b in which the two neutrons
inside their cylinders are absorbed by the target, and
$P_c^{el}P_n^{ab}P_n^{ab}$, that comes from the reaction shown in
Fig.~\ref{fig2}c in which the two halo neutrons are absorbed and the
core is scattered.

In Table~\ref{table1} we indicate which reactions in Fig.~\ref{fig2}
contribute to the various cross sections.  The projectile contains two
neutrons and some of the reactions should be counted twice, since the
symmetric combination also contributes the same amount. The
probabilities shown in Table~\ref{table1} are not explicitly computed.
Instead we compute the cross sections corresponding to each of the
different processes. For example to compute the first process in Table 1
(Fig.~\ref{fig2}a) we note that $P_c^{el}$ corresponds to the probability 
for elastic scattering of the core on the target with the related cross 
section given by eq.(\ref{elas}). The constraints that the two neutrons must 
be outside the cylinders are imposed by removing appropriate parts of 
the initial three-body wave function, see section \ref{sub42}. All other 
processes in Fig.~\ref{fig2}a are analogously computed by replacing the 
first probability by 
the related cross section and removing parts of the three-body wave function 
in the overlap in eq.(\ref{overl}) to account for the other probability 
factors, see section \ref{sub42}. When the probability factors appear 
symmetrically as for processes corresponding to Figs.\ref{fig2}b and 
\ref{fig2}c a specific choice must be made. The details for each of the 
possible cases are described in sections \ref{sub31}, \ref{sub32} and
\ref{sub33}.

Differential cross sections
are given in eqs.(\ref{elas}) and (\ref{abs}), and total cross
sections are obtained after integration over all the variables.
The cross section for a process in which a certain group of constituents 
survive in the final state is then computed by summation of the partial 
cross sections obtained from each of the contributing reactions shown in
Table~\ref{table1}. We emphasize that on top of the contributions in
the last column of Table~\ref{table1} the cross section $\sigma_{nnc}$
also contains the low momentum transfer part due to the Coulomb
interaction that corresponds to the process in Fig.~\ref{fig2}d.

From the cross sections $\sigma_0$, $\sigma_n$, $\sigma_c$, $\sigma_{nn}$,
$\sigma_{nc}$ and $\sigma_{nnc}$ it is easy to obtain several
cross sections of specific interest, as the two neutron removal
cross section ($\sigma_{-2n}$), that is given by all the processes
where the core survives in the final state; the core
removal cross section ($\sigma_{-c}$), given by all the processes
in which the core is destroyed by the target; and the 
interaction cross section ($\sigma_I$), defined as the cross
section given by the reactions where the projectile looses
at least one of its constituents:
\begin{eqnarray} \label{-2nc}
\sigma_{-2n}= \sigma_c + \sigma_{nc} + \sigma_{nnc} \;, \;
\sigma_{-c} = \sigma_0 + \sigma_n + \sigma_{nn}   \;, \; 
\sigma_I    = \sigma_{-2n} + \sigma_{-c} \; .          
\end{eqnarray}

Adding the probabilities shown in Table~\ref{table1} contributing to
each of the cross sections we obtain the probabilities for the
two--neutron removal and core breakup processes as
\begin{eqnarray}
&& P(\sigma_{-2n}) = P_c^{el} + P_n(1-P_c) + P_n(1-P_n)(1-P_c)
\;, \label{P-2n} \\&& P(\sigma_{-c} ) = P_c^{ab} \; . \label{P-c} 
\end{eqnarray}
Therefore the cross sections $\sigma_{-2n}$ and $\sigma_{-c}$ can be
computed as indicated in eq.(\ref{-2nc}), or, alternatively, by
computing the cross sections corresponding to the probabilities in
eqs.(\ref{P-2n}) and (\ref{P-c}).

From eq.(\ref{P-2n}) we see that the two neutron removal cross section
has three contributions. The first one, cross section corresponding to
$P_c^{el}$, is the elastic scattering of the core by the target, and its
differential cross section is given by eq.(\ref{elas}).  Since
$P_c^{el}$ is the probability for the core being inside the cylinder,
only large momentum transfer, $q>q_g$, should be included in
eq.(\ref{elas}).  However, the two--neutron removal cross section also
receives a contribution from the reaction in Fig.~\ref{fig2}d, which
contains the low momentum transfer part ($q_{min}<q<q_g$) of the
core--target interaction.  Thus, $q$ is actually only restricted by
$q>q_{min}$. In addition $\sigma_{-2n}$ also contains the contribution
from the strong interaction between the halo neutrons and the target
with the core as spectator.  These contributions are given by the cross
sections corresponding to the second and third terms in eq.(\ref{P-2n}).
They are computed from eqs.(\ref{elas}) and (\ref{abs}) with
the corresponding optical model neutron-target cross section with the overlap 
in eq.(\ref{overl}) containing the appropriate parts of the three-body wave 
function.  Core breakup cross sections can
be computed, according to eq.(\ref{P-c}), by considering only the
core--target potential and subsequent integration of eq.(\ref{abs}).

\subsection{Two--spectators contributions (Fig.~\ref{fig2}a)}
\label{sub31}

The contributions in Fig.~\ref{fig2}a correspond to processes in which
one halo constituent (participant) interacts with the target, while
the other two (spectators) remain untouched.  They contribute to the
cross sections $\sigma_{nn}$, $\sigma_{nc}$ and $\sigma_{nnc}$ as
shown in the upper part of Table~\ref{table1}.  The general form for
the probability of these reactions is $P_i(1-P_j)(1-P_k)$, where $P_i$
is the probability for the halo constituent $i$ being the participant.

The procedure is then as discussed in Section \ref{sub30}.
The participant--target interaction is described by the potential
eq.(\ref{v0i}) where the nuclear part is given by an optical
potential, that takes into account both absorption and elastic
scattering of the participant by the target. 
The differential cross sections are given by
eqs.(\ref{elas}) and (\ref{abs}), and total cross sections are obtained
after integration over all the variables. These integrations must
take into account the fact that impact parameters for constituents
$j$ and $k$ are larger than the radii of their cylinders.
This is done (see section \ref{sub42}) by removing the appropriate
part of the three--body wave function in eq.(\ref{overl}).

When the core is participant the interference between Coulomb and
nuclear core--target interactions is included in the calculation 
through the differential elastic participant--target cross section
in eq.(\ref{elas}). It is
important to keep in mind that the momentum transfer $q$ in
eq.(\ref{elas}) must be larger than $q_g$ in eq.(\ref{qg}). This
restriction includes only the low impact parameter part of the Coulomb
interaction and allows therefore precisely the processes in question.

\subsection{One--spectator contributions (Fig.~\ref{fig2}b)}
\label{sub32}

The contributions shown in Fig.~\ref{fig2}b correspond to processes in
which two of the halo constituents are inside their cylinders, while
the third one is spectator. In principle each of the two participant
constituents could be either absorbed or scattered by the target, and
therefore the reactions shown in Fig.~\ref{fig2}b can contribute to
the cross sections $\sigma_n$, $\sigma_c$, $\sigma_{nn}$,
$\sigma_{nc}$ and $\sigma_{nnc}$ as seen in the central part of
Table~\ref{table1}.

The probability for occurrence of one of the reactions in
Fig.~\ref{fig2}b is $P_i P_j (1-P_k)$, where $i$ and $j$ refer to the
two participants.  To compute the cross section corresponding to this
reaction we first select the constituent $i$ or $j$ where the
interaction with the target is most conveniently described by the
optical model.  If possible we choose the core.  Then the Coulomb
interaction with the target and the interference between nuclear and
Coulomb interactions are treated carefully.

Cross sections are then computed as a process in which one of the
participants ($i$) interacts with the target through the corresponding
optical potential. Differential cross sections are then given by
eqs.(\ref{elas}) and (\ref{abs}) for elastic scattering and absorption
of constituent $i$ by the target. Again total cross sections are 
computed after integration of (\ref{elas}) and (\ref{abs}) over
all the variables. When the core is participant only the low impact 
parameter part of the Coulomb interaction is allowed, and $q>q_g$ 
in eq.(\ref{elas}).

In this case (reactions in Fig.~\ref{fig2}b) there is a second constituent,
constituent $j$, inside its cylinder that therefore also interacts with 
the target. This fact is included in the calculations by removing the 
appropriate part of the three--body wave function in eq.(\ref{overl}), see
sect.\ref{sub42}. We then ensure that constituent $j$ is close enough to
the target to interact with it. As a result of this interaction constituent
$j$ can be either scattered or absorbed by the target. We then divide
each of the cross sections computed as described above from eqs.(\ref{elas}) 
and (\ref{abs}) in two parts, one of them corresponding
to absorption of constituent $j$ by the target and the other one
corresponding to scattering of constituent $j$. The weight of each of
these parts can be obtained in the optical model for two--body systems,
i.e. simply as $\sigma_{abs}^{(j)}/(\sigma_{abs}^{(j)} + \sigma_{ela}^{(j)})$
for absorption and
$\sigma_{ela}^{(j)}/(\sigma_{abs}^{(j)} + \sigma_{ela}^{(j)})$
for scattering of constituent $j$. Here 
$\sigma_{abs}^{(j)}$ and $\sigma_{ela}^{(j)}$ are
the absorption and elastic cross sections obtained by use of
an optical potential describing the interaction between constituent
$j$ and the target.

Finally, an additional requirement is needed when computing cross sections.
This is due to the fact that particle $k$ is outside its cylinder, and 
therefore not interacting with the target. This is again included in the 
calculation by removing the appropriate part of the three--body wave function
in eq.(\ref{overl}) as described in section \ref{sub42}.

The procedure for constituents $i$, $j$ and $k$ described above permits
to calculate  cross sections for all the processes in Fig.~\ref{fig2}b.
It also permits to compute separately cross sections for processes where
both, constituents $i$ and $j$, are either absorbed or scattered, and also
when one of them is scattered and the other one absorbed.

\subsection{Zero--spectators contribution (Fig.~\ref{fig2}c)}
\label{sub33}

The contributions shown in Fig.~\ref{fig2}c correspond to processes
where all three constituents are inside the corresponding cylinders.
The probability for this reaction is $P_i P_j P_k$, and any of the
three constituents can be either absorbed or scattered by the target.
Thus, this reaction contributes to cross sections where zero, one,
two, or three constituents survive in the final state, as shown in the
lower part of Table~\ref{table1}.

In analogy to the reactions in Fig.~\ref{fig2}b we describe the
core--target interaction with an optical nuclear potential plus the
Coulomb interaction eq.(\ref{v0i}). We then compute cross
sections from eqs.(\ref{elas}) and (\ref{abs}) as a process
in which the interaction between projectile and target takes place 
through the core--target potential. Once more, only the low impact
parameter part of the Coulomb interaction is allowed, and $q>q_g$
in eq.(\ref{elas}).

Now, reactions in Fig.~\ref{fig2}c, constituents $j$ and $k$ are
also inside their cylinders, and therefore both interact with the
target. As before, this is ensured by including in the overlap in
eq.(\ref{overl}) the appropriate part of the three--body wave
function (see sect.\ref{sub42}). Again these two constituents can
be either absorbed or scattered by the target. As in the previous
section, we then divide the computed cross sections given by 
eqs.(\ref{elas}) and (\ref{abs}) in three parts, 
corresponding to absorption of both neutrons, elastic scattering
of both neutrons, and absorption of one of them and scattering
of the other one. The weight of each of these three
contributions is obtained in the optical model for two body systems,
and the part of two absorbed neutrons has a weight given by the
square of $\sigma_{abs}^{(n)}/(\sigma_{abs}^{(n)}+\sigma_{ela}^{(n)})$,
the part of two scattered neutrons has weight given by the square
of $\sigma_{ela}^{(n)}/(\sigma_{abs}^{(n)}+\sigma_{ela}^{(n)})$,
and finally the part of one neutron absorbed and one scattered
has a weight of $2\sigma_{abs}^{(n)}\sigma_{ela}^{(n)}
/(\sigma_{abs}^{(n)}+\sigma_{ela}^{(n)})^2$. Again
$\sigma_{abs}^{(n)}$ and $\sigma_{ela}^{(n)}$ are
the absorption and elastic cross sections obtained by use of
an optical potential describing the neutron--target interaction.

This procedure permits then computations of the cross sections for the
processes in Fig.~\ref{fig2}c where any of the three constituents
can be either absorbed or scattered by the target.

\section{$^6$He and $^{11}$Li on, C, Cu and Pb}
\label{sec4}

In this section we shall apply the method to fragmentation reactions
of the two--neutron halo nuclei $^6$He and $^{11}$Li on three
different targets, one heavy (Pb) with dominating Coulomb contribution
one light (C) with dominating nuclear contribution and an intermediate
mass (Cu) with comparable Coulomb and nuclear contributions. We shall
first briefly give the specifications leading to the initial wave
function. Second we describe how to choose the interactions between
target and halo constituents. The last two subsections contain the
absolute values of the computed cross sections and the momentum
distributions, respectively.

\subsection{Three--body wave functions and intrinsic halo interactions}

The three--body halo wave functions are obtained by solving the
Faddeev equations in coordinate space. This is done by use of the
adiabatic hyperspherical approach. The Faddeev equations are written
in hyperspherical coordinates, and separated into angular and radial
parts. The eigenvalues obtained from the angular part enter into the
radial equation as an effective potential. This procedure has
previously been successfully used, see for instance
\cite{nie00,gar96,gar97}.

The neutron--neutron interaction is given in \cite{gar97}.
In the neutron--core
interaction for 6He we include $s$, $p$ and $d$ waves. The parameters are
adjusted to reproduce the corresponding phase shifts from zero to 20 MeV.
The values are given in \cite{cob98}.

For $^{11}$Li we include $s$ and $p$--waves, and assume spin zero for
the core.  The parameters for the neutron--core interaction 
are adjusted to reproduce the available
experimental data on $^{10}$Li, in particular the presence of a
low--lying $p$--resonance at $500\pm60$ keV with a width of $400\pm60$
keV \cite{you94,abr95,cag99}, and an even lower lying uncertain
virtual $s$--state at $0.15\pm0.15$ MeV \cite{abr95,cag99}.  These
observed states cannot be used directly in computations assuming zero
spin of the $^9$Li-core. The reason is that the spin splitting,
inevitably arising from the finite core spin of 3/2, produces two
levels both of $s$ and $p$-wave character. The measurements refer to
the actual (spin-split) levels. Therefore in the present simplified
model of zero core spin we must aim at statistically averaged $s$ and
$p$-wave energies, which then necessarily must be above the measured
values. We have chosen the realistic model with spin splitting
reproducing the available data and simply reduced the spin splitting
parameter to zero. The more realistic model with finite core spin is
necessary for observables as the two-body invariant mass spectrum, but
in most cases the differences are negligibly small \cite{fed95}.  The
resulting values for the parameters correspond to potential IV in
\cite{gar99}.

It is well known that the use of realistic two--body potentials to
describe a three--body system is leading to three--body binding
energies too small compared to the experimental values. This is a
general problem for few-nucleon systems \cite{carl98}.  For $^6$He and
$^{11}$Li the underbinding is around 0.5 MeV and 150 keV,
respectively. These deviations are substantial compared to the
three-body binding energies, but very small compared to the strengths
of the two-body potentials. To recover the experimental value of the
binding energy we introduce a phenomenological three-body
interaction. The idea is that this three-body force should account for
those polarizations of the particles, which are beyond that described
by the two-body interactions. Thus, this three-body interaction must
be of short range in the hyperradius $\rho$, since it only contributes
when all three particles interact simultaneously.  We choose again the
gaussian radial shape $V_{3b}(r) = V_3 \exp{(-\rho^2/b_3^2)}$, and the
parameters for $^6$He and $^{11}$Li are given in \cite{cob98} and
\cite{gar99}, respectively.

Inclusion of these effective three--body interactions together with
the neutron--neutron and neutron--core interactions described above
give a binding energy and a root mean square radius of 1.0 MeV and
2.50 fm for $^6$He, and 0.3 MeV and 3.35 fm for $^{11}$Li.  These
numbers are for both nuclei consistent with the experimental values,
that are $973.4\pm1.0$ keV \cite{aud95} and $2.57\pm0.10$ fm
\cite{chu89} for $^6$He and $295\pm35$ keV \cite{you93} and
$3.1\pm0.3$ fm \cite{tan92} for $^{11}$Li.  One has
to realize that what is commonly called {\it experimental}
radius for a halo nucleus is deduced from reaction cross section
measurements, and the connection between the few--body wave
function and the cross section is model dependent \cite{alk96}.
The real test is therefore the direct comparison between computed
and experimental reaction cross sections.  The computed $^6$He wave
function corresponds to a probability of roughly 90\% relative
neutron--core $p$--waves and about 10\% $s$--waves, while the
corresponding numbers for $^{11}$Li roughly are 60\% relative
neutron--core $s$--waves and 40\% $p$--waves.

\subsection{Neutron--target and core--target optical potentials}
\label{sub42}

For the neutron--target interaction we use non--relativistic optical
potentials obtained from relativistic potentials through a reduction
of the Dirac equation into a Schr\"{o}dinger--like equation
\cite{udi95}.  These phenomenological potentials in the
Schr\"{o}dinger equation produce the same scattering data as obtained
by use of the relativistic potentials in the Dirac equation
\cite{she86}. In particular, we use the energy--dependent,
$A$--independent parameterizations EDAI-C12 and EDAI-Pb208 in
\cite{coo93} for carbon and lead targets, respectively. For copper
target we use the energy and $A$--dependent potential EDAD--fit1 also
given in \cite{coo93}.  In the three cases the potentials fit the
scattering data in the nucleon energy range from 20 to 1040 MeV.

For the core--nucleus optical potential we use the parameterization
given in \cite{nol87}. The general shape of this potential is:
\begin{equation}
U(r)= - V f(r) - i W g(r) + V_{\mbox{\scriptsize Coul}}
\label{ct}
\end{equation}
with form factors of Woods--Saxon type
\begin{equation}
f(r)=(1+\exp{[(r-r_v A^{1/3})/a_v]})^{-1} \; ,
\label{real}
\end{equation}
\begin{equation}
g(r)=(1+\exp{[(r-r_w A^{1/3})/a_w]})^{-1} \; ,
\label{imag}
\end{equation}
and where $V_{\mbox{\scriptsize Coul}}$ is the Coulomb interaction,
and $A$ the mass number of the target. The energy dependence of the
potential is contained in the parameters $V$ and $W$ as
\begin{eqnarray}
V(A,Z,E)&=&a_0 + a_1 Z A^{-1/3} + a_2 E \; ,
\label{env} \\
W(A,Z,E)&=&b_0 + b_1 Z A^{1/3} + b_2 E \; ,
\label{enw}
\end{eqnarray}
where $Z$ is the charge of the target and $E$ is the energy of the
projectile.

\begin{table}
\begin{center}
\caption{Parameters for the core--target interaction. See eqs.(\ref{ct}) 
-- (\ref{enw}).}
\label{tab2}
\vspace*{0.2cm}
\begin{tabular}{|c|ccc|ccc|}
\hline
Reaction     & $^4$He+C & $^4$He+Cu & $^4$He+Pb &$^9$Li+C & $^9$Li+Cu & $^9$Li+Pb \\ \hline
$r_v$ (fm)   & 1.245    & 1.245     & 1.245     & 1.336   & 1.336     & 1.336   \\
$r_w$ (fm)   & 1.570    & 1.570     & 1.570     & 1.870   & 1.870     & 1.870   \\ \hline
$a_v$ (fm)   & 0.751    & 0.745     & 0.733     & 0.900   & 0.900     & 0.900   \\
$a_w$ (fm)   & 0.651    & 0.635     & 0.607     & 0.750   & 0.750     & 0.750   \\ \hline
$a_0$ (MeV)  & 70       & 110       & 108       & 120     & 220       & 300     \\
$a_1$ (MeV)  & 6.0      & 6.0       & 6.0       & 6.0     & 6.0       & 6.0     \\
$a_2$        & --0.008  & --0.010   & --0.014   & --0.008 & --0.010   & --0.014 \\ \hline
$b_0$ (MeV)  & 17       & 21        & 27        & 28      & 28        & 29      \\
$b_1$ (MeV)  & --1.706  & --1.706   & --1.706   & --1.375 & --1.375   & --1.375 \\
$b_2$        & 0.007    & 0.007     & 0.006     & 0.0025  & 0.0025    & 0.006   \\ \hline
\end{tabular}
\end{center}
\end{table}

The parameters used for the different projectiles and targets are
given in Table~\ref{tab2}. For a $^6$He projectile the radii, $r_v$
and $r_w$, and the diffuseness parameters, $a_v$ and $a_w$, are
directly taken from \cite{nol87}. The strength parameters in
eqs.(\ref{env}) and (\ref{enw}) are also from \cite{nol87} except for
$a_2$ which is reduced to allow for a large energy variation while
simultaneously reproducing the experimental data in \cite{suz94}.  For
a $^{11}$Li projectile the parameters $r_v$, $r_w$, $a_v$ and $a_w$
are from \cite{zah96}, while the parameters in eqs.(\ref{env}) and
(\ref{enw}) are similar to those of $^6$He but slightly modified to
reproduce the available experimental data for $^9$Li--nucleus
scattering \cite{bla93}.

Together with the optical potential we need to specify the value of
the distance $R_{cut}^{(j)}$ which is the limiting impact parameter
determining whether halo constituent $j$ is participant or
spectator. These cutoff parameters play an essential role in the cross
sections.  If the interaction between constituent $i$ and the target
is described by the optical model, the impact parameters for $j$ and
$k$ in eqs.~(\ref{elas}) and (\ref{abs}) must be limited in accordance
with the different geometries in Fig.~\ref{fig2}. This is done in 
practice by including in the overlap eq.(\ref{overl}) only the part of
the initial three--body wave function $\Psi^{(JM)}$ such that the
distances between particles $j$--$i$ and $k$--$i$ are either larger or
smaller than the cutoff distances $R_{cut}^{(j)}$ and $R_{cut}^{(k)}$
depending on the reaction in question. We are therefore forcing particles
$j$ and $k$ to be inside or outside a sphere around particle $i$, instead
of a cylinder along the beam direction. To do the latter is technically
much more difficult, and the use of spheres is a good approximation
providing big advantages from the computational point of view.

Several procedures can be followed to obtain these cutoff parameters:
\begin{itemize}
\item[(a)] Directly determined from the sizes of the constituent and
the target:
\begin{equation}
R_{cut}^{(c)}=r_0 \sqrt{A_t^{2/3}+A_c^{2/3}}, \hspace*{1cm}
R_{cut}^{(n)}=\sqrt{r_0^2 A_t^{2/3}+R_n^2} \; ,
\end{equation}
where the indices $c$ and $n$ refer to the core and the halo neutron,
$A_t$ and $A_c$ are the mass numbers of target and core, and $R_n$ is
the neutron radius. The value of the parameter $r_0$ may vary between
1.1 fm and 1.3 fm.

\item[(b)] In the black disk model the absorption cross section for a
particle $j$ hitting a target is given by $\pi R_{cut}^{(j) 2}$.  We
can then determine the cutoff radius from measured absorption cross
sections. For the projectiles and targets considered in this work some
of these experimental data can be found in \cite{bla93,suz94,sch73}.

\item[(c)] The cutoff radii can also be determined from the measured
mean square radii of target ($\langle r^2 \rangle_t$) and constituent
($\langle r^2 \rangle_j$), which can be found for instance in
\cite{vri87}. The relation to the cutoff radius is then given by
\begin{equation}
\frac{3}{5} R_{cut}^{(j) 2} = \langle r^2 \rangle_t + 
                            \langle r^2 \rangle_j + 2 \;\; \mbox{fm}^2 \; ,
\end{equation}
where 2 fm$^2$ is the square of the range of the nucleon interaction.
This procedure gives values for $R_{cut}^{(j)}$ very similar to 
the impact parameter $b_g$ used in eq.(\ref{qg}) to separate high and low
impact parameters.
\end{itemize}

\begin{table}
\begin{center}
\caption{The cutoff parameters $R_{cut}^{(j)}$ determined according to
the procedures (a), (b) and (c), and the values used in
the actual computations.}
\label{tab3}
\vspace*{0.2cm}
\begin{tabular}{|c|c|ccc|}
\hline
 Target & Method & $R_{cut}^{(n)}$ (fm) & 
           $R_{cut}^{(^4\mbox{\scriptsize He})}$ (fm) & 
           $R_{cut}^{(^9\mbox{\scriptsize Li})}$ (fm)                 \\ \hline
   C    &   (a)  &     2.7--3.1    &      3.4--4.7     &    3.7--5.2  \\
        &   (b)  &     2.6--3.0    &    $\sim$ 4.0     &  $\sim$ 4.7  \\
        &   (c)  &       3.9       &        4.2        &       4.8    \\ 
        &  used  &       3.5       &        4.0        &       4.8    \\  \hline
   Cu   &   (a)  &     4.5--5.3    &      5.2--7.2     &    5.4--7.6  \\
        &   (b)  &     4.9--5.6    &    $\sim$ 6.5     &  $\sim$ 8.5  \\
        &   (c)  &       5.5       &        5.7        &       6.1    \\
        &  used  &       5.4       &        5.7        &       6.1    \\ \hline
   Pb   &   (a)  &     6.6--7.8    &      7.4--10.4    &    7.6--10.6 \\
        &   (b)  &     7.4--7.8    &    $\sim$ 9.0     &  $\sim$ 10.5 \\
        &   (c)  &       7.4       &        7.6        &       7.9    \\
        &  used  &       7.4       &        7.6        &       7.9    \\ \hline
\end{tabular}
\end{center}
\end{table}

These three procedures determine the range of variation for
$R_{cut}^{(j)}$.  The choice of one value or another is a matter of
taste, and within the allowed variations these parameters may be used
for fine tuning the cross sections. The different values obtained are
shown in Table~\ref{tab3} together with the ones actually used in the
calculations.  We see that the range of variation of the cutoff is
rather small for the neutron, and its value consequently is rather
precisely determined. The same happens for $^4$He and $^9$Li on a
carbon target, while for heavier targets like Cu, and especially for
Pb, the cutoff radii vary more.

The adiabatic cutoff energy in eq.(\ref{coulcut}) is chosen as 
$\hbar \omega_a = $ 1 MeV for $^6$He and 0.15 MeV for $^{11}$Li, 
corresponding to the three-body binding energy and half of it, respectively.
This means that $^{11}$Li in the sharp cutoff model breaks up corresponding 
to virtual photon energies down to half of the binding energy.  
The agreement with measured values discussed in the next section 
then indicates a connection with the soft structure revealed in the 
low lying dipole strength \cite{aum99,suz90,cob98,ber93}.
The high energy tail of the virtual photon spectrum must efficiently 
produce breakup of $^{11}$Li \cite{ber88}.

\subsection{Cross sections}

\begin{figure}[t]
\centerline{\psfig{figure=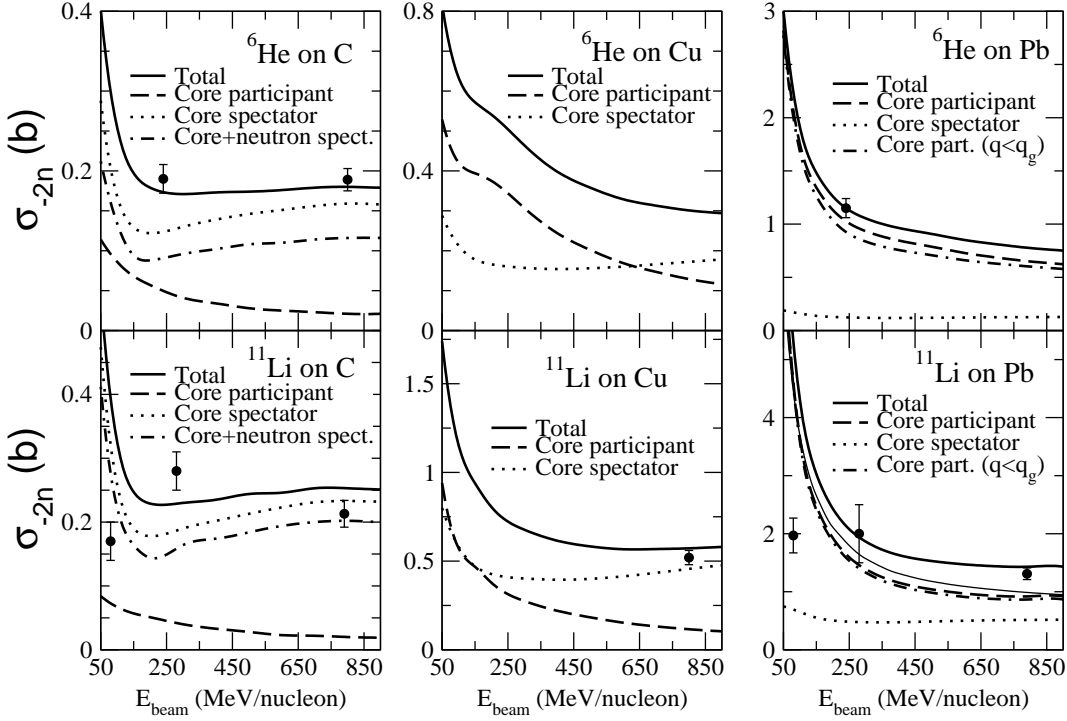,width=13.cm,%
bbllx=2.6cm,bblly=1.5cm,bburx=19cm,bbury=25.5cm,angle=270}}
\vspace*{0.5cm}
\caption[]{Total two--neutron removal cross section, eq.(\ref{-2nc}),
after fragmentation of $^6$He and $^{11}$Li on carbon (left), copper
(middle) and lead (right) as function of beam energy. The long--dashed
and short--dashed lines are the contributions to the total from
participant core and core spectator, respectively. In the carbon case
the dot--dashed line is the contribution assuming that the core and
one of the halo neutrons are spectators. In the lead case the
dot--dashed line is the large impact parameter contribution of the
Coulomb interaction (Fig.~\ref{fig2}d). For $^{11}$Li on Pb the thin
solid line is the total two--neutron removal cross section for 
$\hbar \omega_a$= 0.3 MeV (see text). 
The experimental data are from \cite{tan88,kob89,bla93,zin97,aum99}.}
\label{fig3}
\end{figure}

In Fig.~\ref{fig3} we show two--neutron removal cross sections
obtained after fragmentation of $^6$He and $^{11}$Li on C (left part),
Cu (central part) and Pb (right part). In the three cases the
long--dashed line shows the contribution to the cross section from
elastic scattering of the core by the target (core participant), while
the short--dashed line is the contribution from processes where the
core is spectator. 

For a carbon target the core participant contribution is only
significant at low beam energies, where it contributes by up to 25\%
of the total cross section for the $^6$He projectile, and somewhat
less for $^{11}$Li.  For large beam energies this contribution is at
most 10\% of the total.  The main contribution is given by processes
where the core is a spectator, and only the halo neutrons interact
with the target. This contribution is shown by the short--dashed line
in the figure as computed from the second and third terms in
eq.(\ref{P-2n}). These two terms contain contributions from the
reactions in Fig.~\ref{fig2}a, where one of the neutrons and the core
are spectators, and from the reaction in Fig.~\ref{fig2}b, where only
the core is a spectator.  For comparison we also show in the figure
the two--neutron removal cross section obtained only from the
reactions in Fig.~\ref{fig2}a where one neutron and the core are
spectators (dot--dashed curve).

In the central part we show the results for a Cu target.  Due to the
Coulomb interaction the core participant contribution dominates at low
beam energies, but decreases below that of the core spectator with
increasing energy.  For the $^6$He projectile this crossing appears at
a beam energy of around 600 MeV/nucleon clearly larger than the 100
MeV/nucleon for $^{11}$Li.  Therefore, for energies from 100
MeV/nucleon to 600 MeV/nucleon the main contribution to $\sigma_{-2n}$
comes from core participant reactions for $^6$He projectile and from
core spectator reactions for $^{11}$Li projectile.  In other words, a
Cu target resembles a light target for the $^{11}$Li projectile and a
heavy target for $^6$He.  This is due to the larger size of $^{11}$Li
compared to the one of $^6$He.

The general behavior of $\sigma_{-2n}$ for heavy targets is seen in
the right hand side of Fig.~\ref{fig3}, where the results for a Pb
target are shown.  Due to the large Coulomb interaction the core
participant contribution is clearly dominating, see the dot--dashed
line corresponding to the reaction in Fig.~\ref{fig2}d, where only the
large impact parameter part ($q<q_g$) of the Coulomb interaction is
included.  This contribution is for both projectiles very similar to
the one coming from core participant (dashed), that includes the full
Coulomb core--target interaction, the nuclear core--target
interaction, and the interference between them.  The contribution
given by the neutron--target interaction (core spectator, dotted) is
negligible at low beam energies for both projectiles, while at large
beam energies it contributes around 15\% of the total for $^6$He and
35\% of the total for $^{11}$Li. The agreement with the measured
values is overall rather good considering the incompatibility of some
of the data points. As mentioned at the end of the previous section,
the adiabatic cutoff $\hbar \omega_a$ is chosen to be 0.15 MeV for the 
$^{11}$Li projectile. The result of the expected logarithmic dependence 
on $\hbar \omega_a$ is shown in Fig.~\ref{fig3} for  the two-neutron 
removal cross section for the Pb target where the effect is largest. 
The thin line shows the computed cross section with $\hbar \omega_a$= 0.3
MeV.  The agreement with the experimental points is only slightly worse.
For the C target the minimum value of the momentum transfer is
given by $q_L$ instead of $q_a$ (see sect.~\ref{sec23}), and for Cu target
$q_L$ becomes larger than $q_a$ for beam energies around 150 MeV/nucleon
when $\hbar \omega_a$ = 0.15 MeV and around 400 MeV when
$\hbar \omega_a$ = 0.3 MeV. Therefore the choice of one value or the
other for the adiabatic cutoff energy does not play any role for C target
and insignificant for the Cu target. For $^6$He projectile we used
$\hbar \omega_a$ = 1 MeV, corresponding to its binding energy. 
A variation of this number within a reasonable range produces 
a smaller change than in the case of $^{11}$Li due to its smaller
charge. 

\begin{figure}
\centerline{\psfig{figure=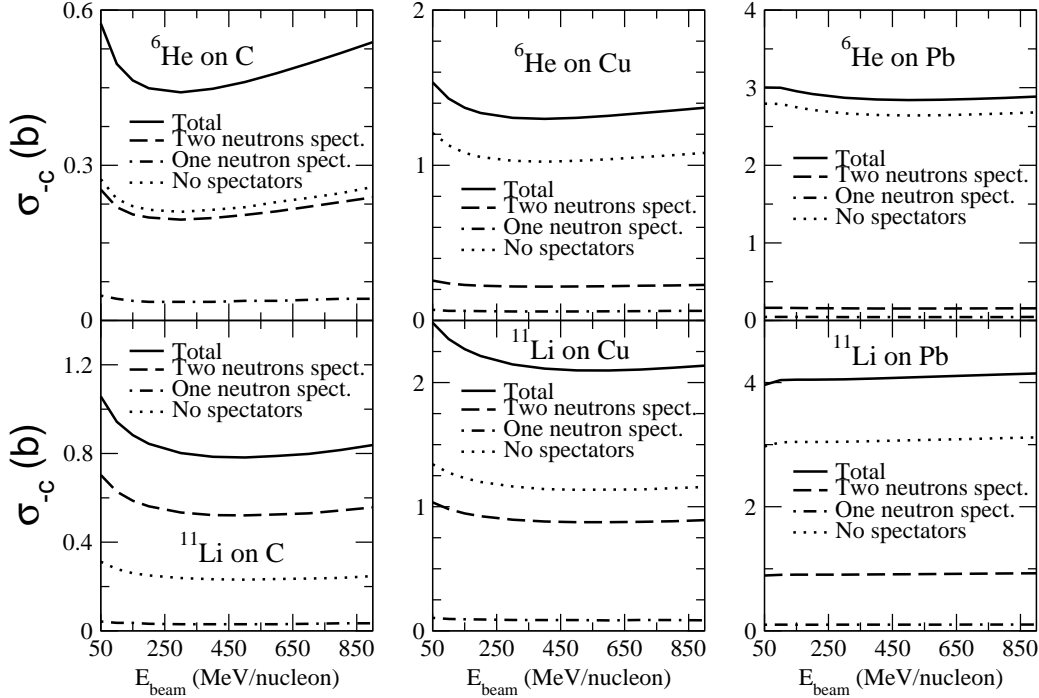,width=13.cm,%
bbllx=2.6cm,bblly=1.5cm,bburx=19cm,bbury=25.5cm,angle=270}}
\vspace*{0.5cm}
\caption[]{Total core breakup cross sections, eq.(\ref{-2nc}), after
fragmentation of $^6$He and $^{11}$Li on carbon (left), copper
(middle) and lead (right) as function of beam energy. The
long--dashed, dot--dashed and short--dashed lines are the
contributions to the total from the reactions in Fig.~\ref{fig2}a (two
neutrons spectators), Fig.~\ref{fig2}b (one neutron spectator) and
Fig.~\ref{fig2}c (no spectators).}
\label{fig4}
\end{figure}

In Fig.~\ref{fig4} we show core breakup cross sections after
fragmentation of $^6$He and $^{11}$Li on C (left), Cu (middle) and Pb
(right). For the three targets we have plotted the contributions from
the reactions in Fig.~\ref{fig2}a, where the core is absorbed while
the two neutrons are spectators (long--dashed line), the contributions
from the reactions in Fig.~\ref{fig2}b where only one of the halo
neutrons is spectator (dot--dashed curve), and the contribution from
the reaction in Fig.~\ref{fig2}c, where there are no spectators
(short--dashed line).

For a carbon target we see that for the $^{11}$Li projectile the
contribution from the two--neutron spectators is dominant, and the
contribution from simultaneous collisions with the target of more than
one constituents is smaller, although still significant.  For the
$^6$He projectile the situation is different, and the no spectators
contribution is even larger than the two--neutron spectators
contribution.  This is a reflection of the difference in size between
these projectiles.  If we denote the neutron--core distance $r_{nc}$
we obtain $\langle r_{nc}^2 \rangle^{1/2}= 4.2$ fm for $^6$He and
$\langle r_{nc}^2 \rangle^{1/2}=5.9$ fm for $^{11}$Li.  In $^6$He the
two halo neutrons are then closer to the core than in $^{11}$Li.
Thus, when the core is absorbed the two halo neutrons have a larger
probability of being spectators for $^{11}$Li than for the $^6$He
projectile.

For a copper target the no spectators contribution is dominant for
core destruction for both projectiles.  Changing from C to Cu we
increase the size of the targets, and therefore the probability for
the halo neutrons being spectators decreases.  Again the smaller size
of $^6$He is responsible for the larger difference between the
long--dashed and short--dashed curves compared to the case of
$^{11}$Li projectile.  The same behavior is observed for a Pb target.
As the target mass increases the probability for the two neutrons
being spectators decreases. The dominant process for a heavy target is
the one where all the three halo constituents interact with the
target.

\begin{figure}
\centerline{\psfig{figure=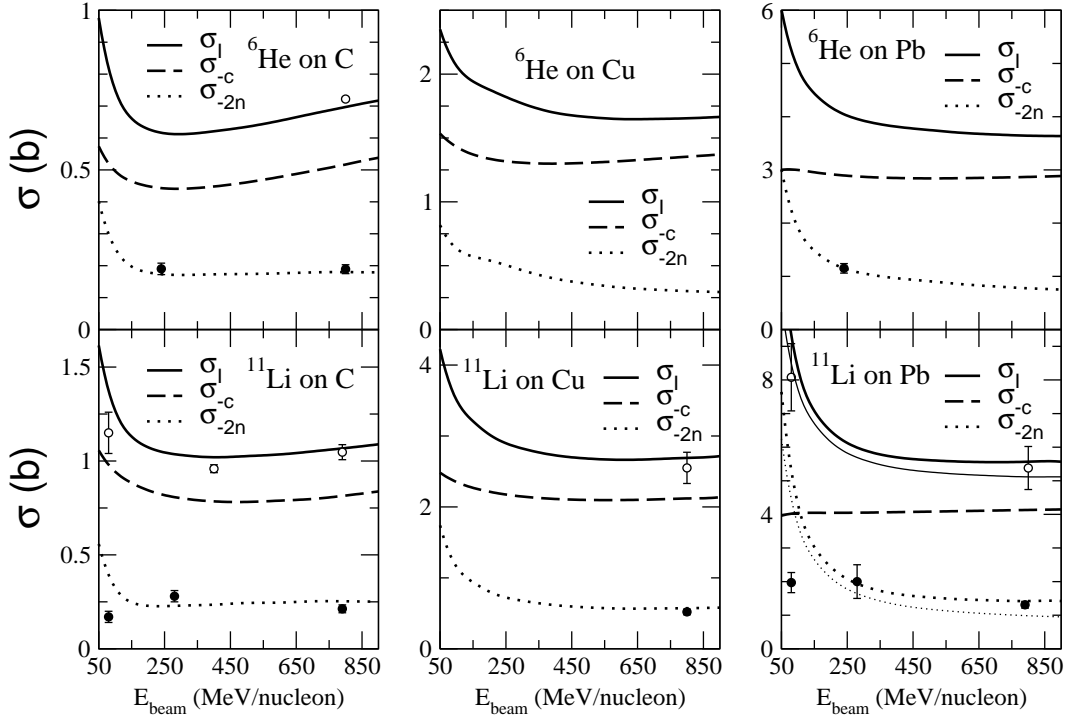,width=13.cm,%
bbllx=2.6cm,bblly=1.5cm,bburx=19cm,bbury=25.5cm,angle=270}}
\vspace*{0.5cm}
\caption[]{Total interaction (solid), two--neutron removal
(long--dashed) and core breakup (short--dashed) cross sections after
fragmentation of $^6$He and $^{11}$Li on carbon (left), copper
(middle) and lead (right). For $^{11}$Li on Pb the thin 
solid and dotted lines are the interaction and two--neutron
removal cross section for $\hbar \omega_a$= 0.3 MeV (see text).
Experimental data are from \cite{tan88,kob89,bla93,zin97,aum99,suz94}.}
\label{fig5}
\end{figure}

In Fig.~\ref{fig5} we plot interaction cross sections for the two
projectiles and the three targets that we are considering. They are
cross sections corresponding to processes where the projectile
undergoes some reaction, and they are given by the sum of the
two--neutron removal and core breakup cross sections.  As a general
rule $\sigma_{-c}$ is larger than $\sigma_{-2n}$, and only for low beam
energies and heavy targets $\sigma_{-2n}$ dominates due to the large
distance Coulomb collisions maintaining the core intact. Again, for
$^{11}$Li on Pb the computed interaction and two-neutron removal
cross sections differ somewhat when the adiabatic cutoff energy 
$\hbar \omega_a$ is increased from 0.15 MeV to 0.30 MeV.

The agreement with the experimental data is good, except for the low
beam energy point for the $^{11}$Li projectile.  However, this
experimental point should be taken with caution, since the
$\sigma_{-2n}$ values for C and Pb targets are not consistent with the
data for $^{11}$Li on beryllium or gold, for which the experimental
values of $\sigma_{-2n}$ at 30 MeV/nucleon are $0.47\pm0.10$ b and
$5.0\pm0.8$ b, respectively \cite{rii92}.

\begin{figure}
\centerline{\psfig{figure=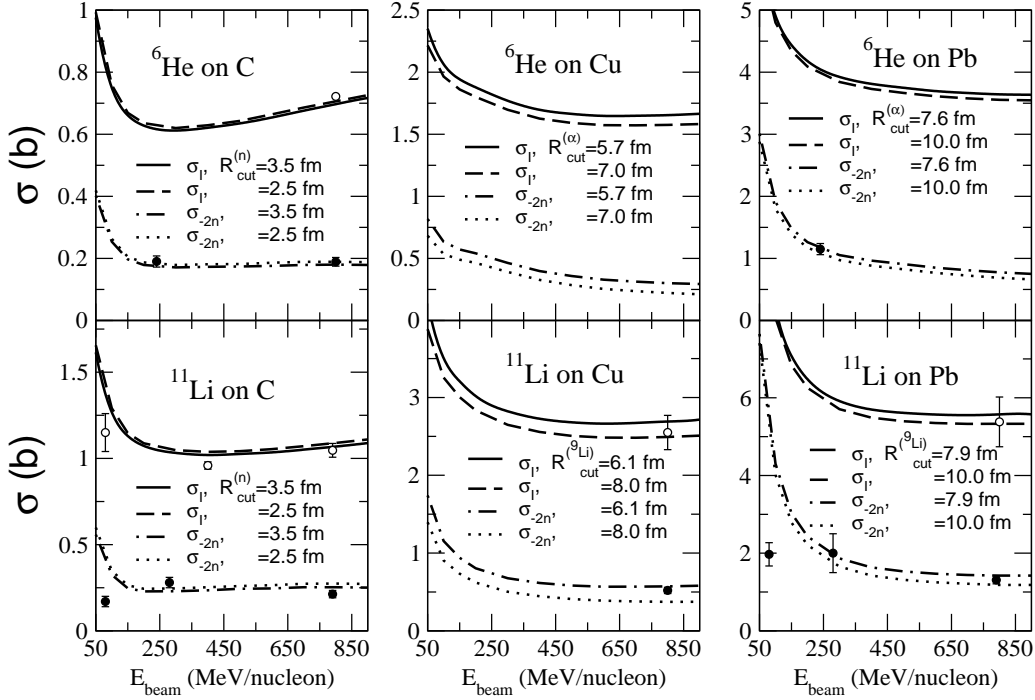,width=13.cm,%
bbllx=2.6cm,bblly=1.5cm,bburx=19cm,bbury=25.5cm,angle=270}}
\vspace*{0.5cm}
\caption[]{Interaction and two--neutron removal cross sections
for $^6$He and $^{11}$Li on C, Cu and Pb for different values of the
cutoff parameters.}
\label{fig6}
\end{figure}

As shown in Table~\ref{tab3} the values of the cutoff parameters
$R_{cut}^{(j)}$ are not uniquely determined.  In Fig.~\ref{fig6} we
examine the dependence of the interaction and two--neutron removal
cross sections on these parameters.  For a carbon target the main
uncertainty appears in the case of a neutron spectator, where the
value of $R_{cut}^{(n)}$ can vary between 2.6 fm and 3.9 fm.  We used
3.5 fm in the previous calculations.  In Fig.~\ref{fig6} we compare
the cross sections with $R_{cut}^{(n)}$=2.5 fm and 3.5 fm, while we
maintain 4.0 fm and 4.8 fm for $R_{cut}^{(^4\mbox{\scriptsize He})}$
and $R_{cut}^{(^9\mbox{\scriptsize Li})}$, respectively.  The
difference between these two calculations is very small.  For Cu and
Pb targets the largest range of variation for the cutoff parameters is
for the core spectator.  In the calculations shown up to now we have
chosen the value obtained with procedure c), but procedures a) and b)
can give significantly larger values for these parameters, see
Table~\ref{tab3}.

In Fig.~\ref{fig6} we also compare $\sigma_I$ and $\sigma_{-2n}$
obtained with $R_{cut}^{(^4\mbox{\scriptsize He})}=$ 5.7~fm and 7.0~fm
and $R_{cut}^{(^9\mbox{\scriptsize Li})} = $ 6.1~fm and 8.0~fm for a
Cu target, and $R_{cut}^{(^4\mbox{\scriptsize He})} = $ 7.6~fm and
10.0~fm and $R_{cut}^{(^9\mbox{\scriptsize Li})} =$ 7.9~fm and 10.0~fm
for a Pb target. It is then striking that even when the difference
between the parameters is substantial, the difference between the
computed cross sections is rather insignificant.  The main reason is
that for these two targets an important contribution comes from the
large impact parameter part of the Coulomb contribution (the reaction
in Fig.~\ref{fig2}d), where the cutoff parameters are unimportant.
This is especially true for the lead target with the large Coulomb
interaction.

The main conclusion obtained from Fig.~\ref{fig6} is that the value
chosen for the cutoff parameters within the range shown in
Table~\ref{tab3} is not substantially modifying the cross sections. We
may view these parameters as relevant for fine tune, while the cross
section sizes and general behavior are maintained.

\subsection{Momentum distributions}

\begin{figure}
\centerline{\psfig{figure=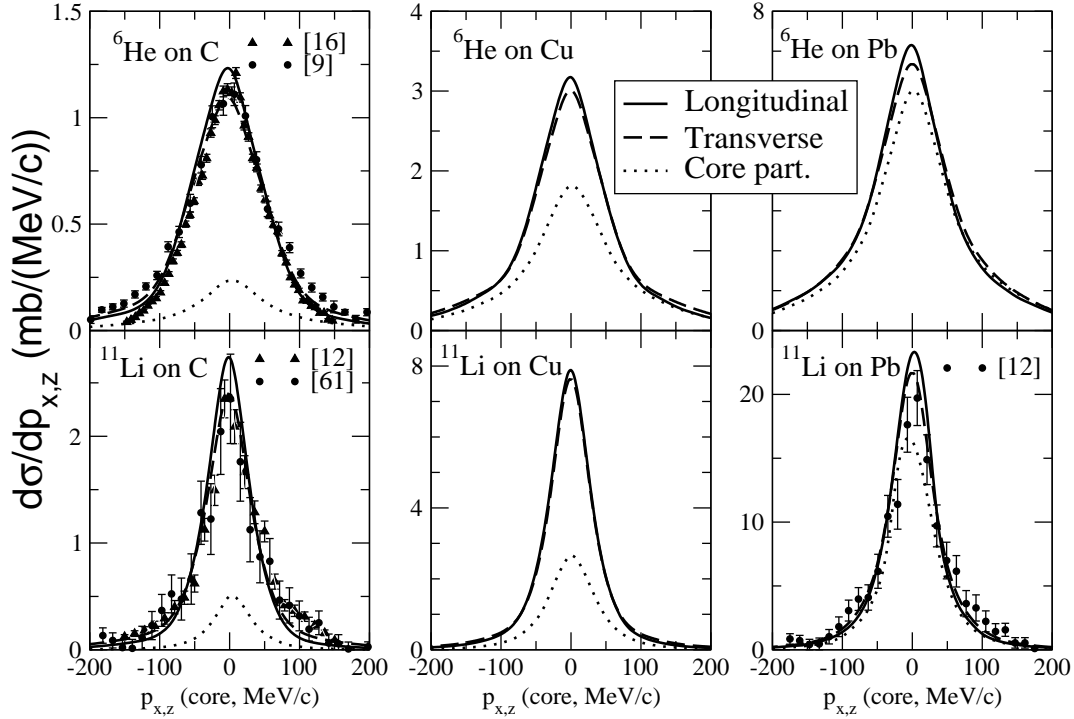,width=13.cm,%
bbllx=2.6cm,bblly=1.5cm,bburx=19cm,bbury=25.5cm,angle=270}}
\vspace*{0.5cm}
\caption[]{Longitudinal (solid) and transverse (long--dashed) core
momentum distributions after fragmentation of $^6$He and $^{11}$Li on
C, Cu and Pb.  The short--dashed line shows the contribution from core
participant.  Experimental data are from \cite{kob92,ale98} for $^6$He
on C, from \cite{hum95,gei} for $^{11}$Li on C and from \cite{hum95}
for $^{11}$Li on Pb. The experimental data have been scaled to the
calculations.}
\label{fig7}
\end{figure}

In Fig.~\ref{fig7} we show the longitudinal and transverse core
momentum distributions after fragmentation of $^6$He and $^{11}$Li on
on C, Cu and Pb. The calculations have been done at a beam energy of 
300 MeV/nucleon.  For this energy the difference between longitudinal 
and transverse momentum distributions is small. The transverse 
distributions are in general wider than the longitudinal, typically 
by 6 to 12 MeV/c.  The experimental data are taken from
\cite{kob92,ale98} for $^6$He on C, and they are obtained at a beam
energy of 400 MeV/nucleon and 240 MeV/nucleon, respectively. For
$^{11}$Li on C the experimental data are taken from \cite{hum95,gei}
and they have been obtained with a beam energy of 280 MeV/nucleon.
For $^{11}$Li on Pb the data are from \cite{hum95} and they correspond
to a beam energy of 280 MeV/nucleon. Within this range of energies the
difference in shape between the computed distributions is not
visible. In all the cases the experimental data are given in arbitrary
units, and they have been scaled to the calculations.  The agreement
between them and the calculations is good.

In Fig.~\ref{fig7} the short--dashed line shows the contribution from
core participant.  For light targets the dominating processes are
those where the core is not participant, and therefore its final
momentum is similar to the one that it had inside the halo projectile
(except for the final neutron--core interaction, that does not modify
significantly the core momentum \cite{gar97b}).  When the target size
increases the weight of the core participant contribution becomes more
important. For Pb this contribution is very close to the total. The
dominating process is then the one shown in Fig.~\ref{fig2}d, where
only the low momentum transfer part of the Coulomb interaction should
be included. The momentum transferred to the core in the collision is
small compared to its initial momentum distribution, which therefore
for a entirely different reason again is left relatively
unchanged.  Although
in this case the three--body continuum wave function should be
used, only the final neutron--neutron interaction has been
included. Nevertheless, due to its large mass, core momentum distributions
are not expected to be strongly modified by the final state interaction
\cite{gar96,gar97b}.  The consequence is that the shape of the core momentum
distributions is almost independent of the target. Only the initial
core momentum distribution within the projectile is decisive.

\begin{figure}
\centerline{\psfig{figure=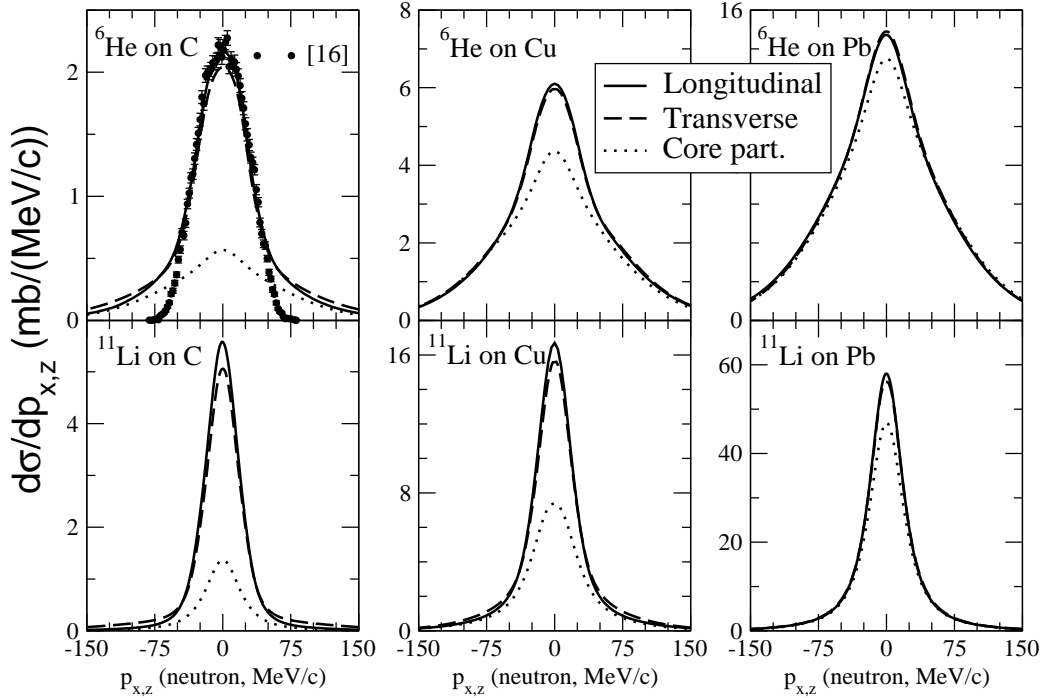,width=13.cm,%
bbllx=2.6cm,bblly=1.5cm,bburx=19cm,bbury=25.5cm,angle=270}}
\vspace*{0.5cm}
\caption[]{The same as in Fig.~\ref{fig7} for neutron momentum 
distributions. The experimental data for $^6$He on C are
from  \cite{ale98}. }
\label{fig8}
\end{figure}

In Fig.~\ref{fig8} we show longitudinal and transverse neutron
momentum distributions after two--neutron removal fragmentation of
$^6$He and $^{11}$Li on C, Cu and Pb. The calculations have been
performed at a beam energy of 300 MeV/nucleon. The difference between
longitudinal and transverse distributions is very small.  Usually the
computed transverse distributions are from 2 to 4 MeV/c wider, and the
difference between them is hardly visible.

As in Fig.~\ref{fig7} the short--dashed curve shows the contribution
from core--participant processes. The behavior is similar to the one
found in Fig.~\ref{fig7}.  For light targets the core spectator
contribution is dominating, while for heavy targets the main
contribution comes from processes where the core participates.  The
only available experimental neutron momentum distribution is for
$^6$He on C \cite{ale98} obtained at a beam energy of 240 MeV/nucleon.
The agreement with the experiment is excellent, and only in the tails
the computed curve is above the experimental distribution. This is due
to the experimental neutron acceptance, that is limited in horizontal
and vertical directions to a momentum of 50 MeV/c.

\begin{figure}
\centerline{\psfig{figure=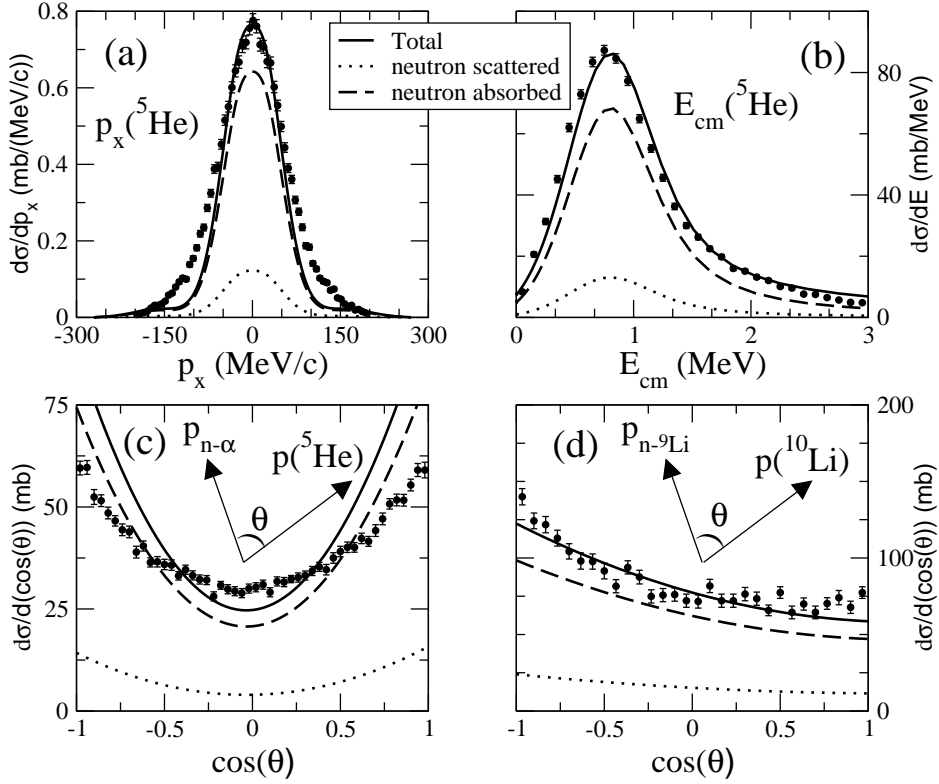,width=13.cm,%
bbllx=2.1cm,bblly=1.cm,bburx=18.5cm,bbury=25.cm,angle=270}}
\vspace*{1.3cm}
\caption[]{Various distributions obtained after fragmentation of $^6$He
and $^{11}$Li on C at 300 MeV/nucleon. 
We only include the dominating parts where
the $\alpha$ particle and one neutron are spectators. (a) is the
transverse distribution of the $^5$He center of mass momentum, (b) is
the invariant mass spectrum of $^5$He, (c) is the angular distribution
with $\theta$ as the angle between the relative $\alpha$--neutron
momentum and the center of mass momentum of $^5$He, and (d) is the
the same as (c) for a $^{11}$Li projectile.
The computed distributions have been convoluted with the
instrumental response \cite{eml99}.  The experimental data are from
\cite{ale98} in (a) and (b), from \cite{chu97} in (c) and from 
\cite{sim99} in (d).}
\label{fig9}
\end{figure}

For a lead target the neutron momentum distribution is dominated by
core participant processes, and in particular by processes where the
momentum transferred to the core is small. As discussed in Section
\ref{sec2}, the final state is properly described for relatively small
impact parameters by the reaction scheme shown in Fig.~\ref{fig1}.
For heavy targets the reaction is dominated by processes where mainly
the low momentum transfer part of the Coulomb interaction
contributes. The breakup process is then relatively gentle, and the
three--body continuum wave function should be used to describe the
final state as shown in Fig.~\ref{fig1a}. Therefore Eqs.(\ref{trans2})
and (\ref{elascoul}) should then be used to describe these processes.

It is well known that neutron momentum distributions are highly
influenced by the final state interaction between fragments, while
this influence is less important for core momentum distributions
\cite{gar96,gar97b} due to the larger mass. In Figs.~\ref{fig7} and
\ref{fig8} we computed all distributions using the reaction scheme of
Fig.~\ref{fig1}, and therefore for the case of Pb target, where the
final three--body continuum wave function should be used, only the final 
neutron-neutron interaction is included. As mentioned this has only a small 
influence on the
core momentum distributions, where the agreement with the experiment
shown in Fig.~\ref{fig7} is satisfactory. However, the computed
neutron momentum distributions after fragmentation on a heavy target
like Pb shown in Fig.~\ref{fig8} are expected to be too broad compared
to the experimental data due to the neglect of the final state
interaction between the neutron and the core for the dominating
process. These computed results, including the neutron-neutron final
state interaction, are still much more realistic than the Fourier
transform of the initial wave function implying that all final state
interactions are neglected.

For completeness we show in Fig.~\ref{fig9} four other interesting
observables obtained after fragmentation of $^6$He and $^{11}$Li.  
As discussed above, inclusion of all the final state interactions 
is required for
heavy targets corresponding to reaction scheme Fig.~\ref{fig1a}. Thus
we focus here only on a C target.  Similar observables for the
$^{11}$Li projectile are already published in \cite{gar99}. 
The calculations are performed for a beam energy of 300 MeV/nucleon,
and the experimental data in Figs.~\ref{fig9}a and \ref{fig9}b from
\cite{ale98} and in Fig.~\ref{fig9}c from \cite{chu97} 
are given in arbitrary units for a beam energy of 240 MeV/nucleon. 
The ones in Fig.~\ref{fig9}d from \cite{sim99} are
also given in arbitrary units for a beam energy of 287 MeV/nucleon.
In ref.\cite{gar98}, where angular distributions for $^6$He projectile
are shown, we inadvertently used a wrong relative phase between interfering 
two-body final states. The conclusions in \cite{gar98} still hold but the 
numerical results now compare slightly better with measured values. 

The calculations in Fig.~\ref{fig9} are in agreement with the experiments, 
except for the angular distribution in Fig.~\ref{fig9}c. The main reason 
for this discrepancy is due to the approximation of spheres instead of the
proper cylinder geometries shown in Figs.~\ref{fig1} and
\ref{fig1a}. The exceptional sensitivity arises in this case, because
the dependence on $\cos \theta$ of the components with neutron-core
relative $p$-wave angular momentum projection $m=\pm 1$ are very flat
while the $m=0$ component is strongly varying \cite{gar98}. Unlike the
other quantities the proper cutoff symmetry is then decisive for this
particular observable. The cylinder would remove relatively more of
the $m=0$ component of the neutron-core $p$-wave and the computation
would approach the measured distribution.
The angular distributions are very different for $^6$He and $^{11}$Li
projectiles as seen by comparing Figs.~\ref{fig9}c and \ref{fig9}d. 
The large average
value for $^{10}$Li reflects the dominating constant contribution from
neutron-core relative $s$-waves and the asymmetric variation is due to the
interference of $s$ and $p$-waves. The strong dependence on the proper
cutoff of the different angular momentum projection states is then not
present. It is especially interesting that the angular distribution for 
$^{10}$Li in Fig.~\ref{fig9}d agrees so well since these computations were 
performed before the measurements were available.

\section{Summary and conclusions}
\label{sec5}

In this paper we formulate a model to describe fragmentation reactions
of two--neutron halo nuclei on any nuclear target, from light to
heavy.  It is well established that nuclear and Coulomb interactions
dominate for light and heavy targets, respectively. Thus both must be
incorporated in a comprehensive description which also should include
intermediate target masses.  We confine ourselves to relatively high
energies where the reaction time can be expected to be short compared
to the time scale of the intrinsic halo motion.

The structure computations proceed in two steps, i.e. few-body models
are used to describe the halo degrees of freedom and then the effects
of the intrinsic structure of the halo constituents are considered.
We adopt the same two-step strategy and start with a reaction model
for weakly bound inert halo particles. Due to the spatial
extension of the initial three-body system the dominating processes
for short-range interactions are independent reactions of each halo
particle (participant) with the target as for free particles. The
participant is then elastically scattered or ``something else'' called
absorption happens while the other two halo particles (spectators)
continue undisturbed under the influence of their mutual interaction.

We use the optical model to describe the participant-target
interaction. Elastic scattering is then described in details while
different processes are considered as absorption which means that all
other channels only remove probability and otherwise
forgotten. However, this is the essence of the information in breakup
experiments where the forward fragmentation products are measured.
Furthermore, absorption in this optical model sense accounts for
intrinsic degrees of freedom related to the target and the halo
constituents. This is in other words consistent with the use of a
three-body model for the halo structure. Going beyond this
approximation in the reaction description requires at least as
accurate a treatment of core degrees of freedom of the initial state.

Another crucial ingredient to understand the reaction mechanisms is
the finite sizes of both halo constituents and target. Conceptually
this is related to the necessity both of accounting for intrinsic
degrees of freedom and for describing more than one simultaneous halo
particle interaction with the target. This becomes clear in the impact
parameter picture where more than one particle at the same time may
arrive at the target position within the interaction cylinder. We
characterize the different processes of absorption and/or elastic
scattering of the different halo particles that simultaneously 
interact with the target. For one of them (the core if possible) we 
employ optical potentials, and for the other(s) we use the part
of the three--body wave function where this constituent is 
close enough to the first one such that it can be considered 
to be inside its cylinder. The model provides absolute cross sections 
for all different processes.

So far we discussed the short-range parts of the interactions. In the
impact parameter picture we may more conveniently formulate this as a
restriction to relatively large momentum transfer to the participant.
Then the reaction mechanism is basically independent halo particle
reactions dictating the division into two independent two-body final
state systems (spectators and participant-target) and in our
formulation resulting in the participant-spectator model. This
division is not for convenience of computation but to include the
proper physics.

The large impact parameters or equivalently the small momentum
transfers are important for long-range interactions. For nuclear halos
this means the Coulomb interaction which is active when the charged
core is a participant. When the target is heavy with a high charge
this gives rise to the main contribution to the breakup cross sections
arising from very distant collisions. The momentum transfer to the
core then excites low lying states of the halo which for Borromean
systems imply the low lying continuum states. These collisions are
gentle and none of the halo particles are absorbed. Consequently the
final state must consist of the three-body halo system and an
independent target. The model distinguishes for these reasons between
small and large momentum transfer and allows therefore simultaneous
treatment of nuclear and Coulomb interactions including interference
terms.

We use the model to investigate fragmentation reactions of $^6$He and
$^{11}$Li on different targets, light as carbon, intermediate as
copper, and heavy as lead. Total interaction, two--neutron removal and
core breakup cross sections are computed as function of beam energy
and found to be in good agreement with the available experimental
data. We compute not only the terms arising from the dominating
reaction mechanism, but also the parts usually neglected: Coulomb
contribution for light targets and nuclear contribution for heavy
targets. Reaction cross sections with targets of intermediate mass and
charge, where nuclear and Coulomb interaction interfere and give
comparable contributions, are also calculated.  The model parameters
are obtained from independent sources like nuclear radii, binding
energies and two-body scattering experiments.

The cross sections for reaching the different final states consisting
of the non-absorbed particles vary with target, projectile and beam
energy. For light targets two neutrons and no core is most probable
while all three halo particles or the core alone is least
probable. For heavy targets absorption of all three particles is most
probable except at low beam energies where simultaneous survival of
all three halo particles is most probable. The smallest probability is
found for core survival alone or the core with one neutron. 

Core destruction is more probable than two-neutron removal for all
energies and targets. Core destruction receives the largest
contribution for heavy targets when all halo particles are absorbed
while for light targets the two-neutron spectator contribution is
comparable in size. Two-neutron removal receives for heavy targets the
largest contributions from core participation while for light targets
the largest contribution arises from the core as a spectator. The
relative sizes of all these cross sections reflect the reaction
mechanisms which in this way are open for experimental tests.

We have also computed core and neutron momentum distributions after
fragmentation. This is a strong test of validity of the model, since
total values obtained after integration could be in good agreement
with experiments even when the non-integrated differential cross
sections differ from measurements.

It is known that final state interactions play an essential role in
the momentum distributions, especially for light particles as
neutrons. For light targets the dominant reactions are the ones with
one or two spectators, and the model accounts properly for the final
state interaction between spectators. The neutron momentum
distribution receives a narrow contribution from one-neutron
absorption and a broader contribution from neutron participation and
subsequent scattering.  The agreement with the available experimental
core and neutron momentum distributions is excellent for both
projectiles.

For heavy targets the low momentum transfer reaction dominates, and
the appropriate reaction mechanism involves the low lying three-body
continuum excitations in the final state. This implies that all three
initial three-body interactions are necessary in the final state. In
the numerical examples we only included two at a time. For core
momentum distributions, where the final state interactions are
expected to be less relevant, we still obtain good agreement with the
experiment. For neutron momentum distributions we include the
neutron-neutron interaction but not the neutron-core interaction. The
distributions are therefore expected to be somewhat too broad.

The model predicts comparable shapes for core and neutron momentum
distributions for light and heavy targets although the reaction
mechanisms are very different. For light targets one-neutron
absorption is the dominating process leaving the other neutron and the
core untouched in the final state under their mutual interaction.  For
heavy targets the dominating process is a distant collision of low
momentum transfer to the core leaving the two neutrons untouched and
the core with a momentum very close to its initial value. Still the
three particles are influenced by the final state interactions. In
both cases the core and the neutrons would appear basically with their
initial momentum distributions modified by final state
interactions. This is not tested in details, but still in qualitative
agreement with experiments.

The model describes a number of different reaction mechanisms and
provides the appropriate relative weights for the corresponding
processes. The same consistent model with one set of parameters is
used throughout for all observables.  In this paper the model is
tested in details by comprehensive calculations for the most studied
two-neutron halo nuclei.  This includes absolute values of the
differential cross sections as functions of beam energy and target.
Dominating contributions to many relative distributions can sometimes
be explained by simpler models since they are surface or tail
dominated. However, the intrinsic degrees of freedom are necessary to
obtain reliable absolute values, especially for less dominating
processes. The details are unnecessary for breakup reactions as long
as predictable total removal probabilities can be obtained as provided
by the phenomenological optical model. 

In conclusion, the model seems to be very successful but the many
predictions should be tested by experiments to tie down in details the
correct reaction mechanisms.

\paragraph*{Acknowledgement.} We thank K. Riisager and E. Nielsen 
for continuous discussions and suggestions.

\end{document}